\documentclass[11pt,3p]{elsarticle}
\usepackage[utf8]{inputenc}
\usepackage[T1]{fontenc}
\usepackage{mathtools,booktabs,amsmath,bm,amssymb,natbib}
\usepackage{subfigure,caption}
\numberwithin{equation}{section}

\newcommand\ddfrac[2]{\ensuremath{\frac{\displaystyle #1}{\displaystyle #2}}}  

\title{Combined searches for dark matter in dwarf spheroidal galaxies observed with the MAGIC telescopes, including new data from Coma Berenices and Draco}

\author[a]{V.~A.~Acciari}
\author[b,al]{S.~Ansoldi}
\author[c]{L.~A.~Antonelli}
\author[d]{A.~Arbet Engels}
\author[e]{M.~Artero}
\author[f]{K.~Asano}
\author[g]{D.~Baack}
\author[h]{A.~Babi\'c}
\author[i]{A.~Baquero}
\author[j]{U.~Barres de Almeida}
% ...

\address[a]{Instituto de Astrof\'isica de Canarias and Dpto. de  Astrof\'isica, Universidad de La Laguna, E-38200, La Laguna, Tenerife, Spain}
\address[b]{Universit\`a di Udine and INFN Trieste, I-33100 Udine, Italy}
\address[c]{National Institute for Astrophysics (INAF), I-00136 Rome, Italy}
\address[d]{ETH Z\"urich, CH-8093 Z\"urich, Switzerland}
\address[e]{Institut de F\'isica d'Altes Energies (IFAE), The Barcelona Institute of Science and Technology (BIST), E-08193 Bellaterra (Barcelona), Spain}
\address[f]{Japanese MAGIC Group: Institute for Cosmic Ray Research (ICRR), The University of Tokyo, Kashiwa, 277-8582 Chiba, Japan}
\address[g]{Technische Universit\"at Dortmund, D-44221 Dortmund, Germany}
\address[h]{Croatian MAGIC Group: University of Zagreb, Faculty of Electrical Engineering and Computing (FER), 10000 Zagreb, Croatia}
\address[i]{IPARCOS Institute and EMFTEL Department, Universidad Complutense de Madrid, E-28040 Madrid, Spain}
\address[j]{Centro Brasileiro de Pesquisas F\'isicas (CBPF), 22290-180 URCA, Rio de Janeiro (RJ), Brazil}
% ...
\address[al]{also at International Center for Relativistic Astrophysics (ICRA), Rome, Italy}
% ...

\cortext[correspondence]{Corresponding authors: C. Maggio, D. Kerszberg, D. Ninci, V. Vitale. Email: contact.magic@mpp.mpg.de}

\date{\today}

\begin{document}

\begin{abstract}
Milky Way dwarf spheroidal galaxies (dSphs) are among the best candidates to search for signals of dark matter annihilation with Imaging Atmospheric Cherenkov Telescopes, given their high mass-to-light ratios and the fact that they are free of astrophysical gamma-ray emitting sources. Since 2011, MAGIC has performed a multi-year observation program in search for Weakly Interacting Massive Particles (WIMPs) in dSphs. Results on the observations of Segue 1 and Ursa Major II dSphs have already been published and include some of the most stringent upper limits (ULs) on the velocity-averaged cross-section $\langle \sigma_{\mathrm{ann}} v \rangle$ of WIMP annihilation from observations of dSphs. In this work, we report on the analyses of 52.1\,h of data of Draco dSph and 49.5\,h of Coma Berenices dSph observed with the MAGIC telescopes in 2018 and in 2019 respectively. No hint of a signal has been detected from either of these targets and new constraints on the $\langle \sigma_{\mathrm{ann}} v \rangle$ of WIMP candidates have been derived. In order to improve the sensitivity of the search and reduce the effect of the systematic uncertainties due to the $J$-factor estimates, we have combined the data of all dSphs observed with the MAGIC telescopes. Using 354.3\,h of dSphs good quality data, 95\,\% CL ULs on $\langle \sigma_{\mathrm{ann}} v \rangle$ have been obtained for 9 annihilation channels. For most of the channels, these results reach values of the order of $10^{-24}\,$cm$^3$/s at ${\sim}1$\,TeV and are the most stringent limits obtained with the MAGIC telescopes so far.
\end{abstract}

\maketitle

\section{Introduction\label{sec:introduction}}

The concept of dark matter (DM) started to gain ground thanks to the work of F.~Zwicky on the galaxies in the Coma galaxy cluster~\cite{1933AcHPh...6..110Z}. After this initial evidence of the existence of DM, several other probes followed, successfully identifying a new, massive, non-luminous, and gravitationally interacting category of matter on galactic, extra-galactic, and  cosmological scales~\cite{2018RvMP...90d5002B}. Among the large number of theories and models that have been proposed along the years to describe its nature~\cite{2018Natur.562...51B}, particle DM within a ${\Lambda}$-Cold Dark Matter (${\Lambda}$CDM) Universe~\cite{1984Natur.311..517B} has been one of the most investigated scenarios. A generic Weakly Interacting Massive Particle (WIMP) is found in super-symmetric extensions of the Standard Model (SM) or extra dimension theories, for instance, and can successfully explain many observational properties of DM on various scales. It is expected to have a mass in the range from a few GeV~\cite{2004JPhG...30..279B} to a few hundred TeV~\cite{PhysRevLett.64.615}, and an interaction cross-section to SM particles typical of the weak scale. Because of their properties and the fact that they are expected to solve the unrelated hierarchy problem, WIMPs have acquired a great popularity in the particular case of \textit{indirect} DM searches.

Depending on the different sensitivities to relevant DM mass ranges, current space-borne gamma-ray telescopes, i.e. Fermi-LAT~\cite{2009ApJ...697.1071A}, ground-based Imaging Atmospheric Cherenkov Telescopes (IACTs), i.e. MAGIC~\cite{2016APh....72...76A}, H.E.S.S.~\cite{2006A&A...457..899A}, and VERITAS~\cite{2015ICRC...34..771P}, and water Cherenkov detectors, i.e. HAWC~\cite{2017ApJ...843...39A}, provide overlapping and complementary results. The characteristic way to indirectly study the nature of DM particles with these detectors is to look for the secondary products of their annihilation or their decay into SM particles. Gamma rays are among the most investigated products because, being stable neutral particles, they can travel straight from their production sites to Earth, thus pointing to their place of origin and giving information about the DM spatial distribution. The most obvious targets where to search for DM are those with high predicted DM densities in the local Universe, such as the Galactic Center and its halo, Galactic DM sub-halos, in which the dwarf spheroidal galaxies (dSphs) of the Milky Way reside, and galaxy clusters~\cite{2017NatPh..13..224C}. When selecting targets of observations, the main points to evaluate are their total expected DM amount and concentration, their distance to Earth, and whether they contain sources of background gamma-ray emission. Due to their high mass-to-light ratio, their proximity to the Earth, and being free of gamma-ray emission from known astrophysical sources, Galactic dSphs are among the most intensively investigated targets. In particular, dSphs are prime targets for IACTs as the extension of their DM halos\footnote{Please note that we use dSph in the following to denote the DM halo of dSphs.} is typically of the order of the field of view of the telescopes, a fact that simplifies the analysis with respect to that for more extended sources such as e.g. the Galactic halo. Additionally, their existence as a part of the population of Galactic DM sub-halos is clearly predicted by the ${\Lambda}$CDM hierarchical structure formation scenario. Despite numerous observation campaigns and sophisticated analyses, no hint of DM signatures has been observed from these targets so far and only constraints on DM particle cross-section have been set~\cite{2020Galax...8...25R}.

MAGIC has observed various Milky Way dSphs in search for a DM signal since the very beginning of the telescopes' operation. In this paper, the latest individual and combined results of the indirect DM search program in dSphs performed by MAGIC are presented. The concept of indirect DM searches with IACTs is introduced in Section~\ref{sec:gamma-ray-signal}, followed in Section~\ref{sec:MAGIC-dSphs} by a description of the MAGIC telescopes and the dSphs data samples considered for this work. The details of the selection and low level treatment of the data from the newly observed Draco and Coma Berenices dSphs are presented in Section~\ref{sec:dSphs-analysis}. The high-level DM analysis, the so-called \textit{full likelihood} analysis, is described in Section~\ref{sec:likelihood-analysis}. The individual results of this analysis for the Draco and Coma Berenices dSphs are presented in Section~\ref{sec:Draco-and-Coma-results}, while in Section~\ref{sec:Ursa-and-Segue-results} the results from Ursa Major~II and Segue~1 dSphs are recalled. The combined analysis and subsequent limits are shown in Section~\ref{sec:combined-limits}, followed by a discussion and a comparison with previous results provided by MAGIC and other experiments. In Section~\ref{sec:conclusions}, the content of the paper is summarized and conclusions are drawn.

\section{Gamma-ray signal from annihilating DM\label{sec:gamma-ray-signal}}

Indirect DM searches with IACTs aim at detecting gamma-ray fluxes produced by the annihilation or decay of WIMPs in regions of the sky where a sizeable concentration of DM is expected, the so-called {DM over-densities}. The differential gamma-ray flux, integrated in a certain aperture $\Delta\Omega$, can be expressed as the product of a \textit{Particle Physics} (\textit{PP}) factor and an \textit{Astrophysical} (or $J$-) factor. In case of DM annihilation, it can be written as:
\begin{equation}
    \ddfrac{d\Phi(\Delta\Omega)}{dE}=\ddfrac{1}{4\pi} \ddfrac{\langle\sigma_{\mathrm{ann}} v \rangle}{2m^{2}_{\mathrm{DM}}} \ddfrac{dN}{dE} \times J(\Delta\Omega).
    \label{eq:gamma-ray-flux}
\end{equation}
The first three factors on the right hand side of the equation compose the PP-factor. It contains all the information regarding the DM model under consideration: the DM particle mass $m_{\mathrm{DM}}$, the gamma-ray spectrum $\ddfrac{dN}{dE} = \sum\limits_{i=1}^{n} \mathrm{Br}_{i}\ddfrac{dN_{i}}{dE}$ produced per annihilation in $n$ possible channels and weighted by the corresponding branching ratios $\mathrm{Br}_i$, and the velocity-averaged annihilation cross-section $\langle\sigma_{\mathrm{ann}} v \rangle$. This last quantity is the one that is either measured (in case of detection of a DM signal) or constrained (in case of a non-detection) in indirect DM annihilation searches.

Whereas the \textit{PP} term is determined only by the nature of DM, and hence is the same for every source\footnote{This is true under the assumption that there is only one kind of DM particle, or that the relative abundance of more than one kind of DM particle is the same in all investigated targets, which is not necessarily the case. Nevertheless, this assumption is reasonable until a DM signal detection allows us to investigate it.}, the $J$-factor $J(\Delta\Omega)$ incorporates the specific source's DM distribution and its distance from the observer. It is expressed as the integral of the DM density ($\rho$) squared along the line-of-sight (l.o.s.) and over the solid angle $\Delta\Omega$ under which the target is observed:
\begin{equation}
    J(\Delta\Omega)=\int_{\Delta\Omega}d\Omega'\int_{\mathrm{l.o.s.}}dl \rho^{2}(l,\Omega').
    \label{eq:J-factor}
\end{equation}
In this work, the spectral and morphological templates for the gamma-ray emission of the observed dSphs were, thus, estimated from the gamma-ray spectra expected from WIMP annihilation and the estimated DM distribution, following Equation~\ref{eq:gamma-ray-flux}. DM particles in the mass range 0.07--100\,TeV and annihilating into the SM particle pairs $e^+e^-$, $\mu^+\mu^-$, $\tau^+\tau^-$, $W^+W^-$, $ZZ$, $HH$, $b\bar{b}$, $t\bar{t}$ and $\gamma \gamma$ have been considered.
The expected average gamma-ray spectrum per annihilation process $d N_{i}/d E$ was taken from~\cite{2011JCAP...03..051C}, while the emission morphology of the source has been modeled with the $J$-factor differential values, i.e. the $J$-factor distribution with respect to the angular distance from the center of the target, provided in~\cite{2015ApJ...801...74G} (see Section~\ref{sec:MAGIC-dSphs} for details).

\section{The MAGIC telescopes and the dSphs data samples\label{sec:MAGIC-dSphs}}

The Florian G\"{o}bel Major Atmospheric Gamma Imaging Cherenkov (MAGIC) telescopes are a system of two 17\,m diameter IACTs operated in coincidence in the so-called stereoscopic mode. The telescopes are located at the Observatorio del Roque de los Muchachos (ORM) on the Canary Island of La Palma (Spain), at an altitude of ${\sim}2200$\,m above sea level. Thanks to their large reflector surfaces, the new trigger systems~\cite{Strom:2020yei} and the wide alt-azimuth movement, the MAGIC telescopes can detect gamma rays in the energy interval ranging from ${\sim}30$\,GeV to ${\sim}100$\,TeV~\cite{2020A&A...635A.158M}, with an angular resolution of ${\sim}0.08^\circ$ at the 68\,\% containment radius of the point spread function for energies above 200\,GeV~\cite{2016APh....72...76A}.

After the observation campaigns on Segue~1~\cite{2014JCAP...02..008A} and Ursa Major~II~\cite{2018JCAP...03..009A} dSphs, MAGIC started a new multi-year DM program for the study of two additional dSphs, namely, Draco and Coma Berenices. The classical dSph Draco was discovered in 1954 by the Palomar Observatory Sky Survey~\cite{1955PASP...67...27W}. Coma Berenices dSph belongs to the so-called ultrafaint dSphs discovered in the Sloan Digital Sky Survey~\cite{2007ApJ...654..897B} in 2006. Both targets have been observed with the MAGIC telescopes for ${\sim}50$\,h each, for a total scheduled amount of ${\sim}100$\,h. The observations were carried out in wobble mode~\cite{1994APh.....2..137F} and only one pair of wobble positions was adopted, which reduces systematic differences in the acceptance of signal (also called ON) and background-control (also called OFF) regions~\cite{2014JCAP...02..008A}.

The aim of the project was to enlarge and diversify the pool of dSphs observed with the MAGIC telescopes, with the goal of increasing the chances of detection of a DM signal from new unexplored regions, of mitigating the effect of the systematic uncertainties related to the expected DM content of the selected target and, in case of no detection, of improving the constraints on $\langle \sigma_{\mathrm{ann}} v \rangle$. The targets considered for the observations were selected among the dSphs presented in~\cite{2015ApJ...801...74G}. The selection criteria combined observability from the MAGIC site, as large as possible estimated $J$-factor values and as small as possible related statistical uncertainties. Table~\ref{tab:dSphs-list} presents several relevant quantities for the two newly selected targets, Draco and Coma Berenices dSphs, and for the two ultrafaint dSphs previously observed by MAGIC, Segue~1 and Ursa Major~II (which are part of the combined analysis). The MAGIC observations on Triangulum~II~\cite{2020PDU....2800529A} have not been included in this list, nor in the subsequent combined analysis, due to the present uncertainty on the dynamical equilibrium of the object~\cite{2017ApJ...838...83K} and, thus, the lower reliability of its $J$-factor estimate.

\begin{table}
    \centering
    \caption{List of the dSphs investigated in the MAGIC multi-year dSph DM project. For each dSph, we report: the logarithm of its total $J$-factor and its respective uncertainty, the maximum angular distance $\theta_\mathrm{max}$ and the one containing 50\,\% of the assumed DM emission $\theta_\mathrm{0.5}$ (i.e. $J(\theta_{0.5}) = 0.5 \times J(\theta_{\mathrm{max}})$) taken from~\cite{2015ApJ...801...74G}, as well as the effective observation time $T_{\mathrm{eff}}$ and the year of data taking by MAGIC. The maximum angular distance is the angular distance of the outermost member star used to evaluate the velocity dispersion profile. It coincides with the most conservative truncation radius of the assumed DM annihilation emission.}
    \begin{tabular}{cccccc}
        \textbf{Target}   
        & \textbf{\vtop{\hbox{\strut \boldmath$\log_{10} J(\mathrm{\theta_\mathrm{max}})$}\hbox{\strut {\bf [GeV}\boldmath$^{2}${\bf cm}\boldmath$^{-5}]$}}}    
        & \textbf{\vtop{\hbox{\strut \boldmath$\theta_\mathrm{max}$} \hbox{\strut {\bf [deg]}}}} & \textbf{\vtop{\hbox{\strut \boldmath$\theta_\mathrm{0.5}$} \hbox{\strut {\bf [deg]}}}} & \textbf{\vtop{\hbox{\strut \boldmath$T_{\mathrm{eff}}$} \hbox{\strut {\bf ~[h]}}}} & \textbf{Year} \\ 
        \toprule
        Coma Berenices  & $19.02^{+0.37}_{-0.41}$ & 0.31 & $0.16^{+0.02}_{-0.05}$ & 49.5 & 2019  \\
        Draco           & $19.05^{+0.22}_{-0.21}$ & 1.30 & $0.40^{+0.16}_{-0.15}$ & 52.1 & 2018 \\
        Ursa Major~II   & $19.42^{+0.44}_{-0.42}$ & 0.53 & $0.24^{+0.06}_{-0.11}$ & 94.8 & 2016--2017\\
        Segue~1         & $19.36^{+0.32}_{-0.35}$ & 0.35 & $0.13^{+0.05}_{-0.07}$ & 157.9 & 2011--2013\\
        \bottomrule 
    \end{tabular}
    \label{tab:dSphs-list}
\end{table}

\begin{figure}[t]
    \begin{subfigure}
        \centering
        \includegraphics[width=0.49\textwidth]{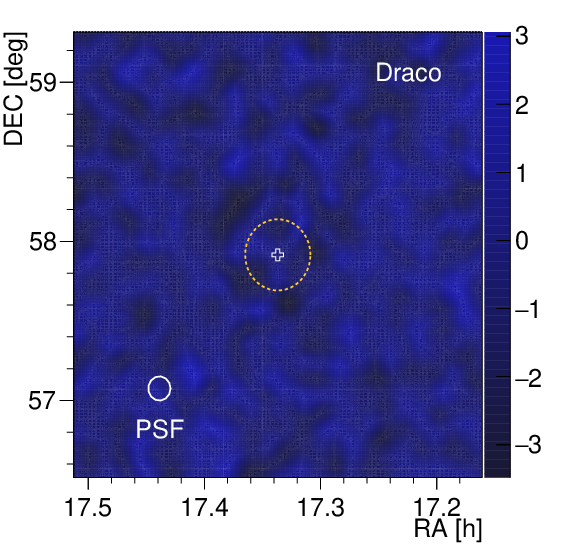}
    \end{subfigure}
    \begin{subfigure}
        \centering
        \includegraphics[width=0.49\textwidth]{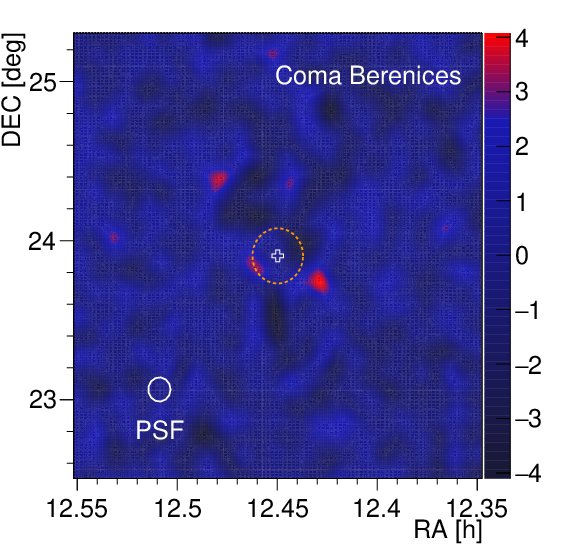}
    \end{subfigure}
    \caption{Significance skymaps in the Draco (left) and Coma Berenices (right) dSphs field of view, respectively. They have been produced with a test statistic (see Equation~17 of~\cite{1983ApJ272317L}), applied on a smoothed and modelled background estimation. The color scale on the right side of each figure represents the test statistic value distribution. The empty white cross refers to the center of the target, and the orange dashed circle delimits the signal region defined in this analysis, here corresponding to an optimized $\theta^2$ cut of $0.05\,\deg^2$ and $0.03\,\deg^2$ for Draco and Coma Berenices dSphs, respectively (see Section~\ref{sec:likelihood-analysis} for the details on the optimization). The white solid circle on the bottom left corner of each figure shows the MAGIC point spread function.}
    \label{fig:dSphs-skymaps}
\end{figure}

As shown in Table~\ref{tab:dSphs-list}, both previously and newly selected dSphs present $J$-factor values above $10^{19}$\,GeV$^2$cm$^{-5}$. Note that for Ursa Major~II, the previous analysis~\cite{2018JCAP...03..009A} already used the $J$-factor value from~\cite{2015ApJ...801...74G}, while for the analysis of Segue~1~\cite{2014JCAP...02..008A} a previous $J$-factor estimation was adopted~\cite{2014JCAP...02..008A,2010PhRvD..82l3503E}. Also, the limits on the WIMP annihilation cross-section obtained on Segue~1 in~\cite{2014JCAP...02..008A} were later used in a combined analysis with Fermi-LAT results~\cite{2016JCAP...02..039M}. In the latter work, another different $J$-factor was used, namely the value used by Fermi-LAT previously in~\cite{2015PhRvL.115w1301A}. Since the study by~\cite{2015ApJ...801...74G}, the DM content of Segue~1 has been re-evaluated multiple times~\cite{2015PhRvL.115w1301A,2015MNRAS.453..849B,2016MNRAS.461.2914H,2016MNRAS.462..223B,2017MNRAS.466..669C}, with the value from~\cite{2015ApJ...801...74G} agreeing with most of the more recent estimates. Hence we adopt the result from~\cite{2015ApJ...801...74G} also for Segue~1 for this work. The data sample of Segue~1 dSph is the same as the one presented in~\cite{2014JCAP...02..008A}, except for a light revision of the analysis that will be described in Section~\ref{sec:Ursa-and-Segue-results}. On the other hand, the data sample of Ursa Major~II dSph is the same as in~\cite{2018JCAP...03..009A}. In the following section, the results of the low level analysis performed on the data taken for Draco and Coma Berenices dSphs are presented.

\section{MAGIC low level analysis of Draco and Coma Berenices dSphs data\label{sec:dSphs-analysis}}

Draco dSph is the target with the third-highest $J$-factor after Segue~1 and Ursa Major~II in the list from Section~\ref{sec:MAGIC-dSphs}. It was observed with the MAGIC telescopes from March to September 2018 in the zenith angle range between $29^\circ$ and $46^\circ$. Starting from June 30, a degraded mirror reflectance caused a lower data quality that had to be taken into account. The dataset was hence divided in two samples and paired to specific Monte Carlo (MC) simulations, better reflecting the status of the instrument for each sample. The observations of Coma Berenices dSph were performed from the end of January to the beginning of June 2019. The target was observed at low zenith angles, between $5^\circ$ and $36^\circ$, and, as no major changes affected the instrument during that period, one set of MC simulations was sufficient for this analysis.

Data from Draco and Coma Berenices dSphs were reduced using the standard MAGIC analysis software MARS~\cite{2013ICRC...33.2937Z}. As it is usually the case in DM studies, excellent data quality was required in order to guarantee good performances and non-coherent systematic uncertainties (the effect of the coherent ones is negligibly small) in the estimation of the residual background below 1.5\,\%, as evaluated in~\cite{2016APh....72...76A}. Thus, strict data selection criteria have been applied, especially with respect to atmospheric conditions for which we required more than 85\,\% transmission~\cite{Fruck:2014mja}. It resulted in the selection of 52.1\,h and 49.5\,h of excellent quality data for Draco and Coma Berenices dSphs, respectively.

The particle identification was carried out using a Random Forest event classification method~\cite{2008NIMPA.588..424A} that assigns a parameter, called \textit{hadronness}, to each event. After data reduction, no significant gamma-ray excess over the background was detected in either of the field of views, as shown in Figure~\ref{fig:dSphs-skymaps} for Draco (left) and Coma Berenices (right) dSphs.

\section{Likelihood method for high-level DM analysis\label{sec:likelihood-analysis}}

Once the events are reconstructed and tagged as gamma-ray candidates, the observed numbers of events as a function of the reconstructed energy are fitted by a likelihood with the signal intensity as free parameter, whose value is estimated using standard likelihood maximization. The likelihood analysis is binned (in reconstructed energy), which allows a better treatment of the systematic uncertainty of the irreducible background with respect to an unbinned analysis~\cite{2018JCAP...03..009A}. The binned likelihood function $\mathcal{L}$, whose parameter of interest is the weighted-averaged annihilation cross-section $\langle \sigma_{\mathrm{ann}} v \rangle$, for each target $t$, each dataset $\bm{\mathcal{D}}$ corresponding to a different set $k$ of Instrument Response Functions (IRFs), and for each pointing direction $i$ is written as follows (removing the indexes $t, k$ and $i$ from the right hand side of the equation to avoid overloading it):
\begin{align}
    \begin{split}
        \mathcal{L}_{tki}(\langle \sigma_{\mathrm{ann}} v \rangle;\bm{\nu}|\bm{\mathcal{D}}) & = \mathcal{L}(\langle\sigma_{\mathrm{ann}} v \rangle;J,\{b_{j}\}_{j=1,...,N_{\mathrm{bins}}},\tau | (N_{\mathrm{ON},j},N_{\mathrm{OFF},j})_{j=1,...,N_{\mathrm{bins}}}) \\ 
         & = \prod_{j=1}^{N_\mathrm{bins}}\left[ \ddfrac{(g_{j}(\langle\sigma_{\mathrm{ann}} v \rangle,J)+b_{j})^{N_{\mathrm{ON},j}}}{N_{\mathrm{ON},j}!} e^{-(g_{j}(\langle\sigma_{\mathrm{ann}} v \rangle,J)+b_{j})} \times
        \ddfrac{(\tau b_{j})^{N_{\mathrm{OFF},j}}}{N_{\mathrm{OFF},j}!}e^{-\tau b_{j}} \right] \\ 
        &~~~~\times \mathcal{T}(\tau|\tau_{\mathrm{obs}}, \sigma_{\tau}) \\
        &~~~~\times \mathcal{J}(J|\log_{10} J_{\mathrm{obs}},\sigma_{\log_{10} J}),
    \end{split}
    \label{eq:binned-likelihood-formula}
\end{align}
where $j$ runs over the number of bins in energy $N_{\mathrm{bins}}$. In Equation~\ref{eq:binned-likelihood-formula}, $\bm{\mathcal{\nu}}$ represents the nuisance parameters which are the $J$-factor $J$, the expected number of background events $b_{j}$ and the OFF/ON acceptance ratio $\tau$. The likelihood function is then written as the product of three terms. The first one consists of Poissonian functions for the number of observed events in the ON region ($N_{\mathrm{ON},j}$), i.e. the region from which the signal is extracted, and the number of observed events in the OFF region ($N_{\mathrm{OFF},j}$), i.e. the region used to evaluate the background. The second one ($\mathcal{T}$) corresponds to the likelihood for the OFF/ON acceptance ratio, parametrized by a Gaussian function with mean $\tau_{\mathrm{obs}}$, computed as the ratio of the number of the observed events in regions adjacent to the OFF and ON ones, and variance $\sigma^2_{\tau}$ which includes both statistical and systematic uncertainties following $\sigma_{\tau}=\sqrt{\sigma_{\tau,\mathrm{stat}}^2 + \sigma_{\tau,\mathrm{syst}}^2}$, where  $\sigma_{\tau,\mathrm{syst}}=1.5\,\%\cdot\tau$ as estimated in~\cite{2016APh....72...76A}. It is important to note that $\tau$ does not depend on the energy bin $j$, it is hence considered as a global nuisance parameter. The third term ($\mathcal{J}$) is the likelihood function for the logarithm of the $J$-factor, also parametrized by a Gaussian function with mean $\log_{10} J_{\mathrm{obs}}$ and variance $\sigma^2_{\log_{10}J}$. In this analysis, the statistical uncertainty on the $J$-factor, here treated as a nuisance parameter, dominates over other systematic uncertainties. Therefore no additional systematic in the gamma-ray efficiency is considered in the analysis, in particular regarding the derivation of the upper limits, where an additional systematic uncertainty of ${\sim}30$\,\% on the effective area is usually considered for gamma-ray sources (see e.g.~\cite{2013A&A...549A..23A}).

The expected number of gamma-rays $g_{j}$ depends on the free parameter $\langle \sigma_{\mathrm{ann}} v \rangle$, that is the parameter of interest, and on the $J$-factor nuisance parameter as follows:
\begin{equation}
    g_{j}(\langle \sigma_{\mathrm{ann}} v \rangle, J) = T_{\mathrm{obs}}\int_{E'_{\mathrm{min},j}}^{E'_{\mathrm{max},j}} dE' \int_0^{\infty}dE\ddfrac{d\phi(\langle \sigma_{\mathrm{ann}} v \rangle, J)}{dE} A_{\mathrm{eff}}(E)G(E'|E)
    \label{eq:number-of-expected-gamma-rays}
\end{equation}
where $T_{\mathrm{obs}}$ is the total observation time, the extremes of the integral $E'_{\mathrm{min},j}$ and $E'_{\mathrm{max},j}$ are respectively the minimum and the maximum energies of the $j$-th energy bin, $A_{\mathrm{eff}}$ is the effective area and $G$ is the probability density function for the energy estimator $E'$, given the true energy $E$. The latter probability density function, together with the effective area, represent the IRFs. They are computed starting from MC simulations of diffuse gamma rays that follow the spatial distribution of the expected DM-induced signal of each dSph. The morphology of each dSph was modeled using the \textit{Donut} MC method~\cite{2018JCAP...03..009A}. Following the procedure in~\cite{2016JCAP...02..039M}, the estimator $g$ is restricted within the physical region (e.g. $g \geq 0$) during the likelihood maximization procedure. To combine the results obtained for each dSph, the final likelihood function is then written as the product:
\begin{equation}
    \mathcal{L}(\langle \sigma_{\mathrm{ann}} v \rangle;\bm{\nu}|\bm{\mathcal{D}}) = \prod_{t=1}^{N_{\mathrm{target}}} \prod_{k=1}^{N_t} \prod_{i=1}^{2} \mathcal{L}_{tki}(\langle\sigma_{\mathrm{ann}} v \rangle;\bm{\nu_{tki}}|\bm{\mathcal{D}_{tki}}).
    \label{eq:combined-likelihood-formula}
\end{equation}
where $t=1 \ldots 4$ identifies the four targets considered in this work (see Table~\ref{tab:dSphs-list}), $k$ varies from 1 to $N_t$ where $N_t$ is the number hardware configuration expressed by the IRFs under which the target $t$ was observed ($N_t$ equals 2 and 1 for Draco and Coma Berenices dSphs respectively, see Section~\ref{sec:dSphs-analysis}, while $N_t$ equals 4 for Segue~1~\cite{2014JCAP...02..008A} and 1 for Ursa Major~II~\cite{2018JCAP...03..009A}), and the index $i$=1,~2 denotes the two pointing directions.
The estimation of $\langle \sigma_{\mathrm{ann}} v \rangle$ is performed using the profile likelihood ratio test $\lambda_p$, defined as:
\begin{equation}
    \lambda_p(\langle \sigma_{\mathrm{ann}} v \rangle|\bm{\mathcal{D}})=\ddfrac{\mathcal{L}(\langle \sigma_{\mathrm{ann}} v \rangle;\bm{\hat{\hat{\nu}}}|\bm{\mathcal{D}})}{\mathcal{L}(\widehat{\langle \sigma_{\mathrm{ann}} v \rangle};\bm{\hat{\nu}}|\bm{\mathcal{D}})},
    \label{eq:profile-likelihood-ratio}
\end{equation}
where $\bm{\hat{\nu}}$ and $\widehat{\langle \sigma_{\mathrm{ann}} v \rangle}$ are the values that maximize the likelihood, and $\bm{\hat{\hat{\nu}}}$ maximize the likelihood for a fixed value of $\langle \sigma_{\mathrm{ann}} v \rangle$. Making use of Wilks' theorem, the one-sided upper limits (ULs) on the velocity-averaged cross-section at the 95\,\% confidence level (CL) are obtained when $\lambda_p$ fulfills the following constraint\footnote{Because of the degeneracy between $\langle \sigma_{\mathrm{ann}} v \rangle$ and $J$ in the gamma-ray flux computation (see Equation~\ref{eq:gamma-ray-flux}), and the fact that $J$ is considered as a nuisance parameter with log-normal probability density function, the coverage of our confidence intervals is not exactly 95\,\%. Using simulations, we have verified that the recipe in Equation~\ref{eq:upper-limits} produces an over-coverage. We nevertheless computed the ULs using this rule since over-coverage produces conservative limits and in order to be able to perform meaningful comparisons with previous results and those from other experiments using the same prescription (see, e.g.~\cite{2015PhRvL.115w1301A,2017PhRvD..95h2001A,2020PhRvD.101j3001A,2020PhRvD.102f2001A}).}
:
\begin{equation}
    -2\ln \lambda_p(\langle \sigma_{\mathrm{ann}} v \rangle^{\mathrm{UL}}|\bm{\mathcal{D}}) = 2.71.
    \label{eq:upper-limits}
\end{equation}
The sensitivity to $\langle \sigma_{\mathrm{ann}} v \rangle$, defined as the average UL that would be obtained by an ensemble of experiments in the case of no DM signal (i.e. the null hypothesis $\langle \sigma_{\mathrm{ann}} v \rangle = 0$), can be approximated by:
\begin{equation}
    \langle \sigma_{\mathrm{ann}} v \rangle_{\mathrm{sensivity}}=\langle \sigma_{\mathrm{ann}} v \rangle^{\mathrm{UL}}-\widehat{\langle \sigma_{\mathrm{ann}} v \rangle}.
    \label{eq:sensitivity}
\end{equation}
This definition of the sensitivity is independent of the actual limit value and is therefore used to optimize the analysis cuts without introducing any bias to the final result. The optimization is done from fast simulations of the null hypothesis for the cuts on the \textit{hadronness} and the squared angular distance $\theta^{2}$ between the nominal position of the target and the reconstructed event direction. For this calculation, the parameter $g$ is not restricted to only positive values and $J$ is considered with no uncertainty. These energy-dependent optimized cuts have then been applied blindly to the data, as described in~\cite{2018JCAP...03..009A}. All the likelihood functions reported in this section are implemented in the open source tool gLike~\cite{javier_rico_2021_4601451}, which provides the joint likelihood maximization as a function of $\langle \sigma_{\mathrm{ann}} v \rangle$, as well as the profiling over the nuisance parameters. The combined limit was also cross-checked using the independent software package LklCom~\cite{tjark_miener_2021_4597500}.

\begin{figure}[ht!]
    \begin{subfigure}
        \centering
        \includegraphics[width=.49\textwidth]{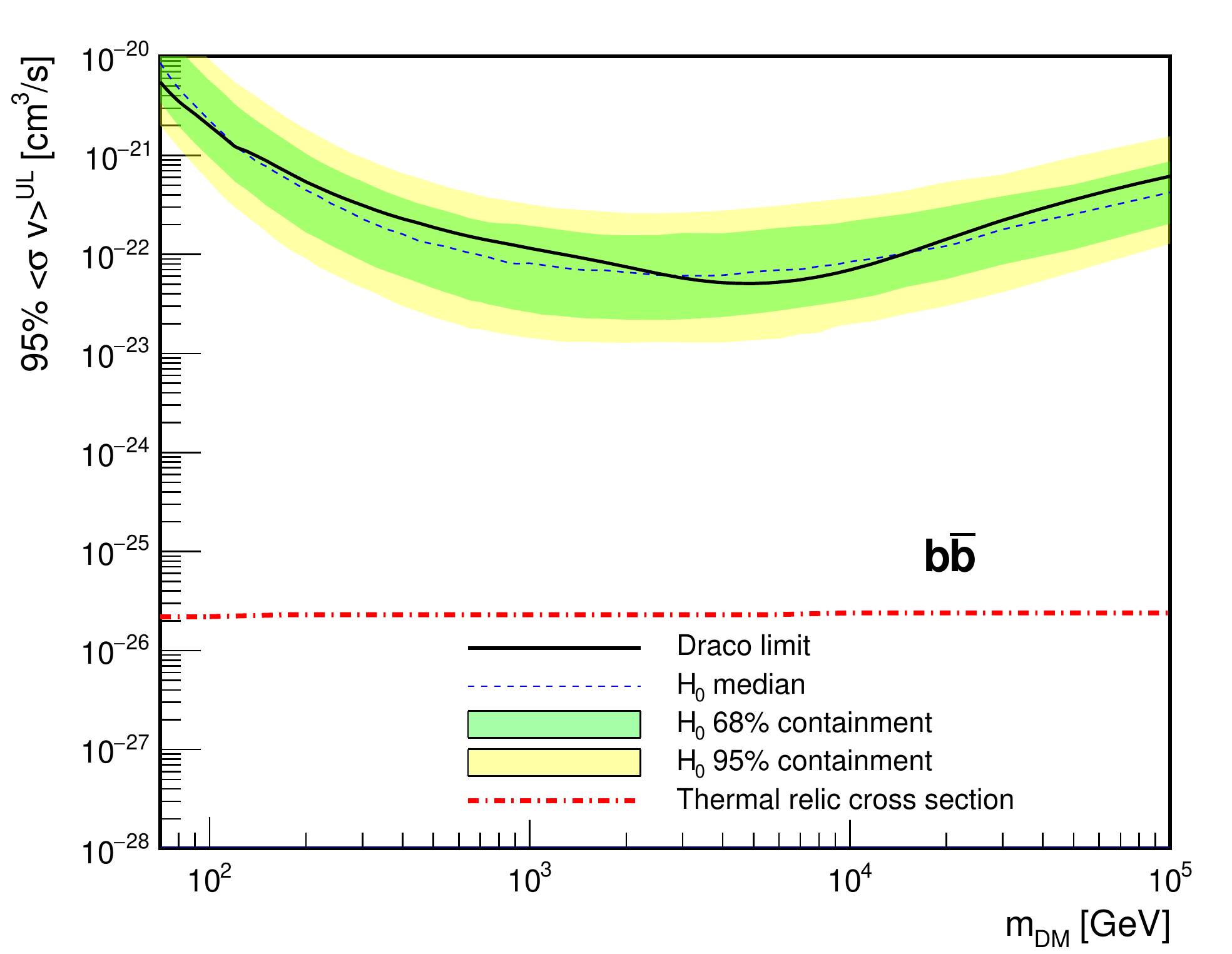}
    \end{subfigure}
    \begin{subfigure}
        \centering
        \includegraphics[width=.49\textwidth]{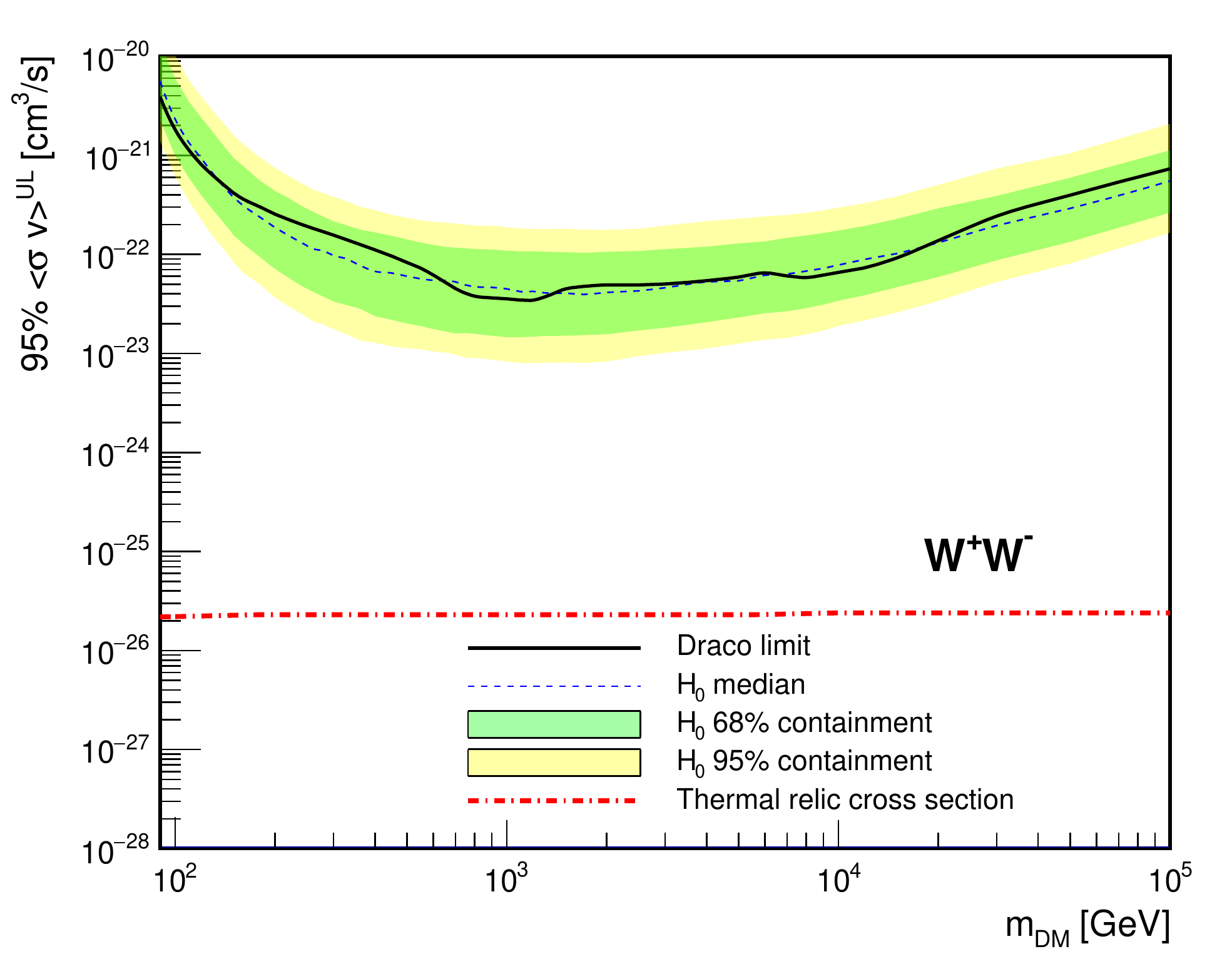}
    \end{subfigure}
    \begin{subfigure}
        \centering
        \includegraphics[width=.49\textwidth]{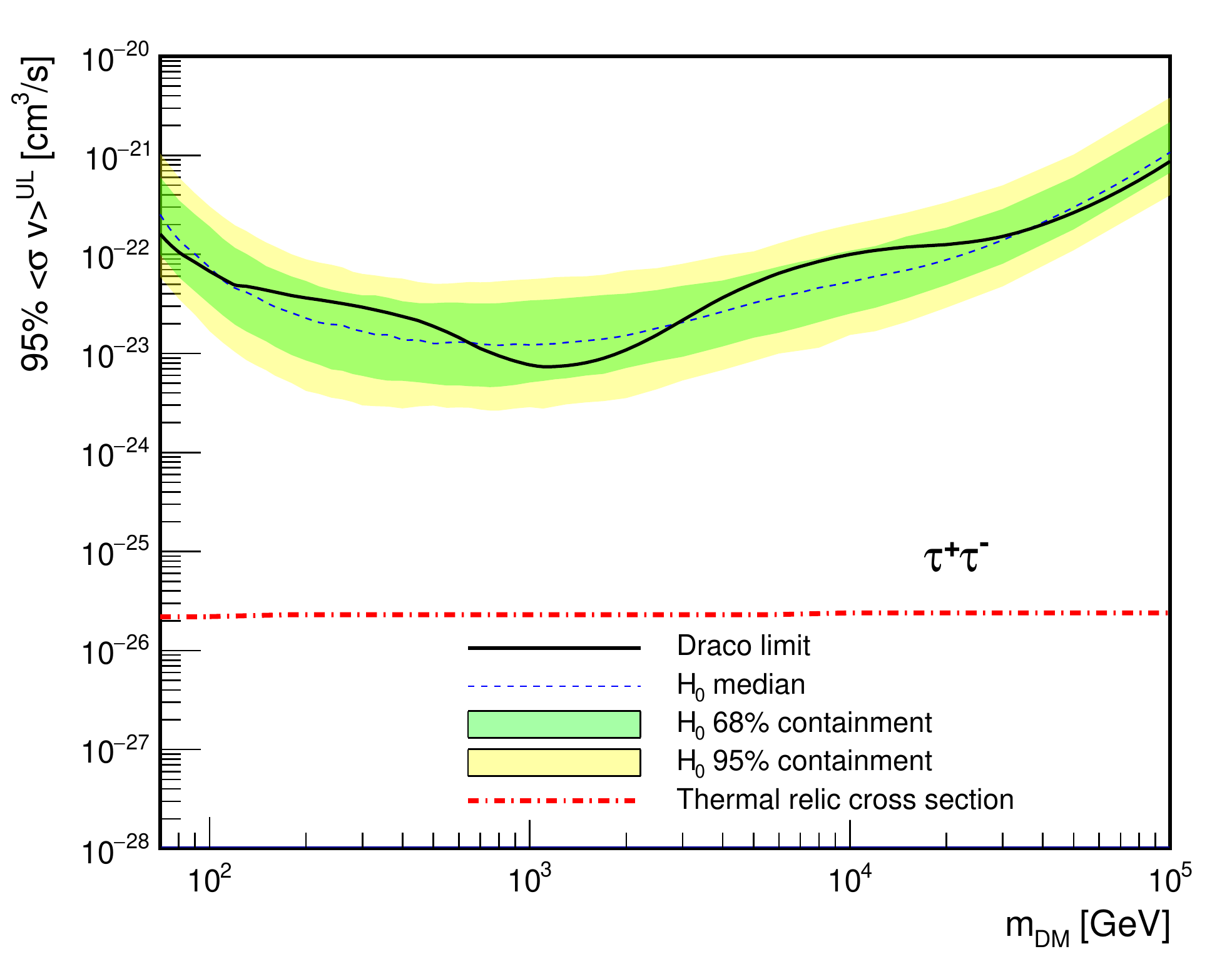}
    \end{subfigure}
    \begin{subfigure}
        \centering
        \includegraphics[width=.49\textwidth]{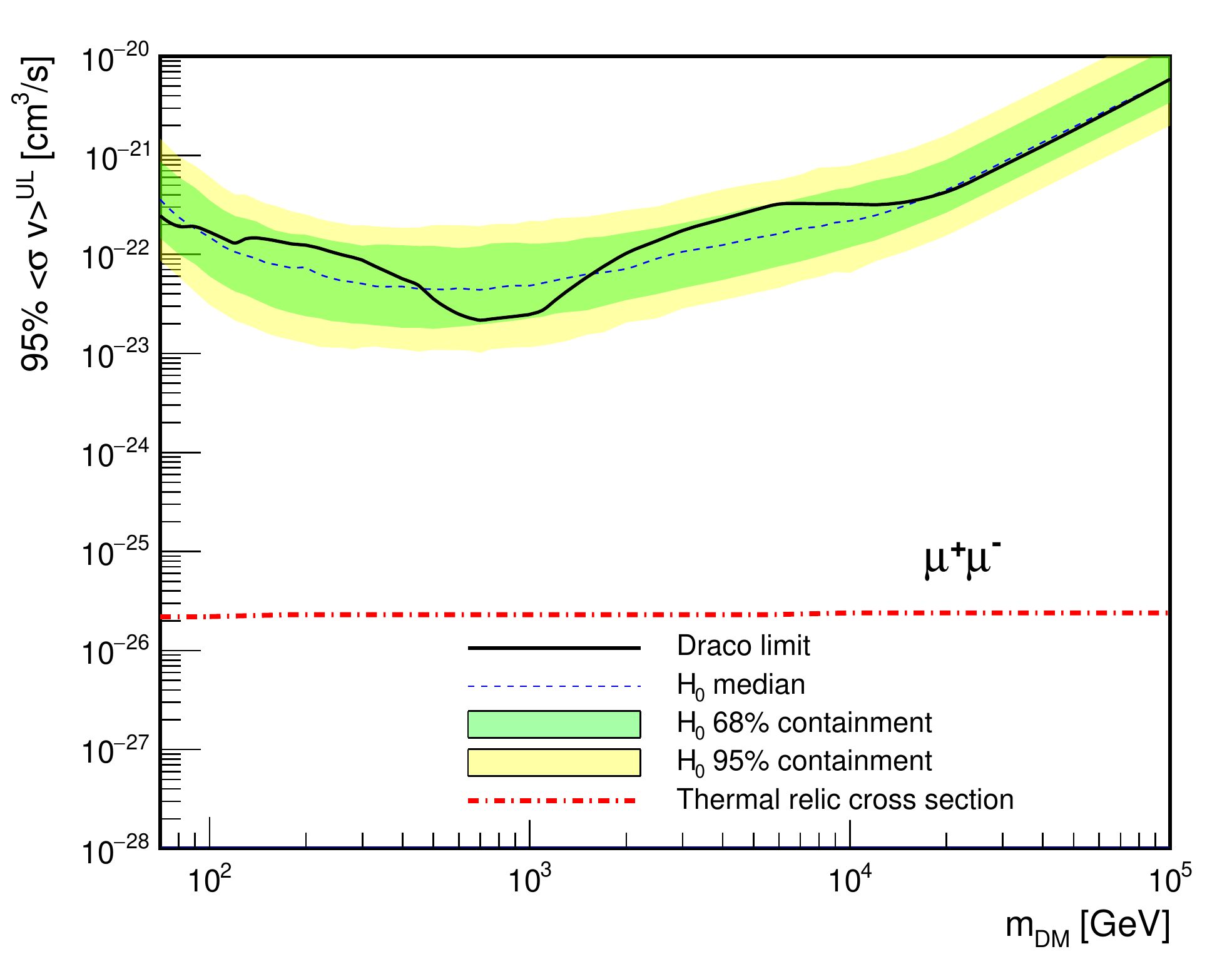}
    \end{subfigure}
    \caption{95\,\% CL ULs for $\langle \sigma_{\mathrm{ann}} v \rangle$ for DM annihilation into $b\bar{b}$, $W^{+}W^{-}$, $\tau^{+}\tau^{-}$ and $\mu^{+}\mu^{-}$ pairs, as representative annihilation channels for both leptonic and hadronic  interactions. The black solid line indicates the observed limits obtained for 52.1\,h of Draco dSph observations, while the blue dashed line is the median of the 300 realizations of the null hypothesis. The green and yellow bands represent the two sided 68\,\% and 95\,\% containment bands respectively. The red dashed line shows the thermal relic cross-section~\cite{2012PhRvD..86b3506S}.}
    \label{fig:limits-Draco}
\end{figure}

\section{DM results from Draco and Coma Berenices dSphs\label{sec:Draco-and-Coma-results}}

\begin{figure}[ht!]
    \begin{subfigure}
    \centering
        \includegraphics[width=0.49\textwidth]{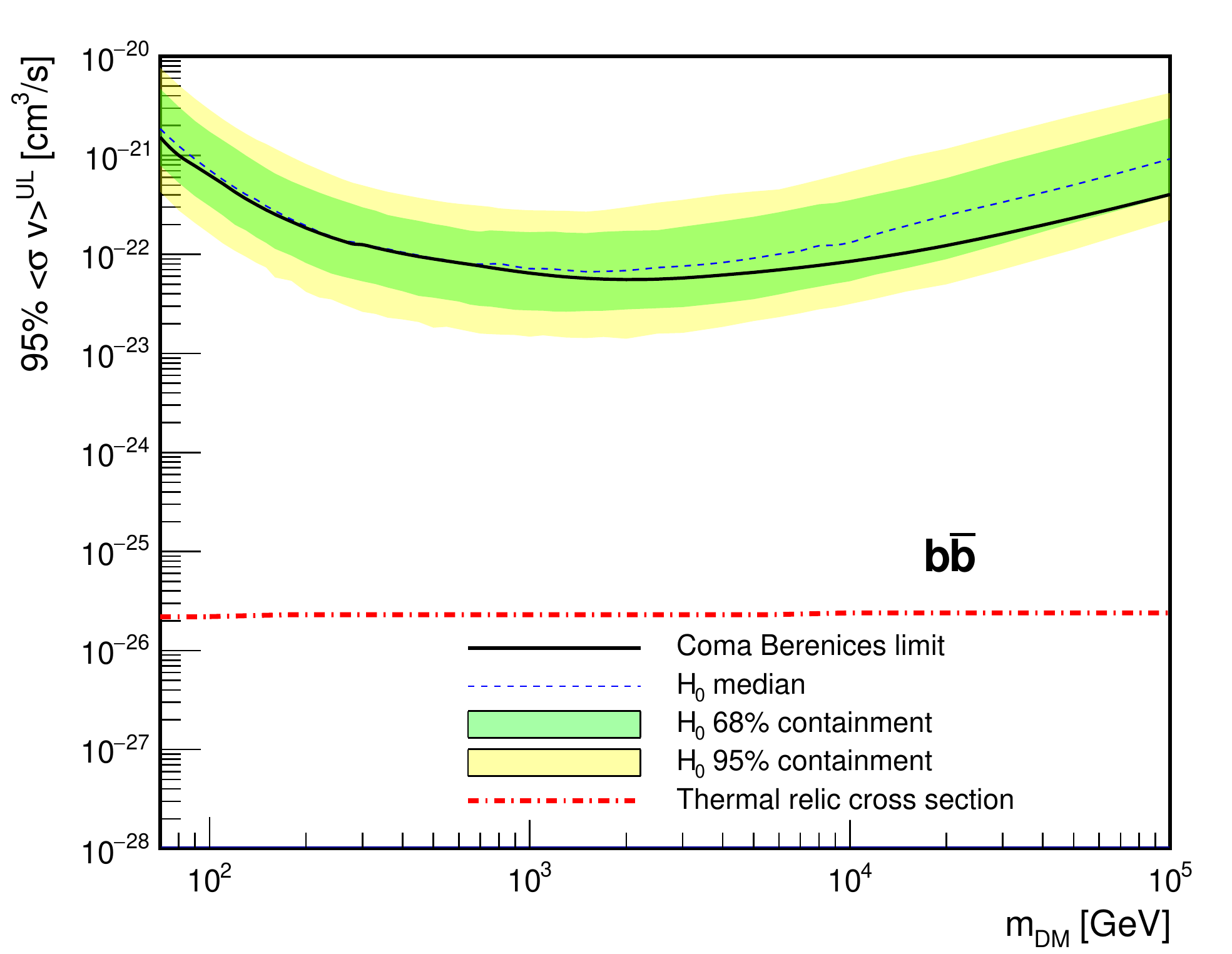}
    \end{subfigure}
    \begin{subfigure}
        \centering
        \includegraphics[width=0.49\textwidth]{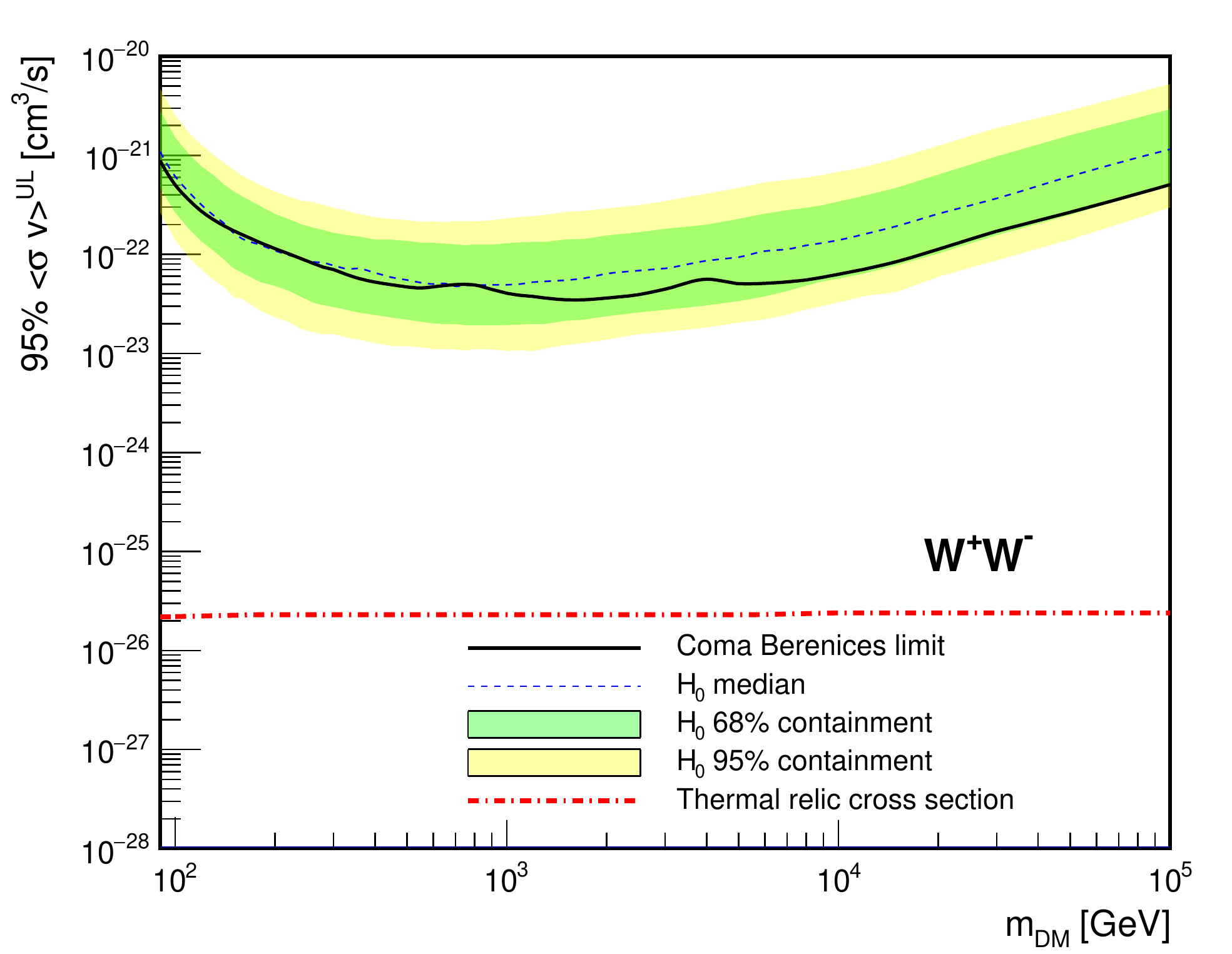}
    \end{subfigure}
    \begin{subfigure}
        \centering
        \includegraphics[width=0.49\textwidth]{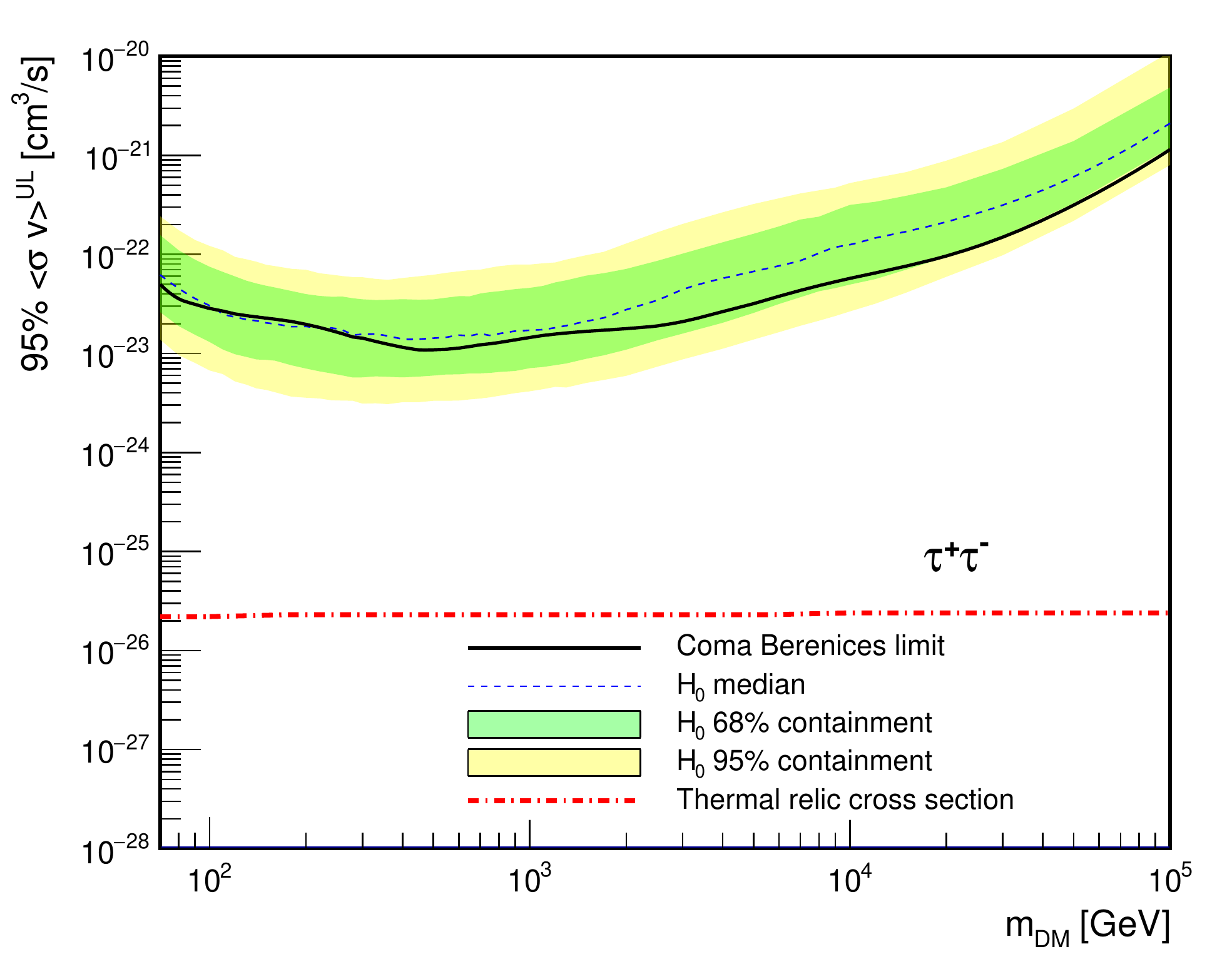}
    \end{subfigure}
    \begin{subfigure}
        \centering
        \includegraphics[width=0.49\textwidth]{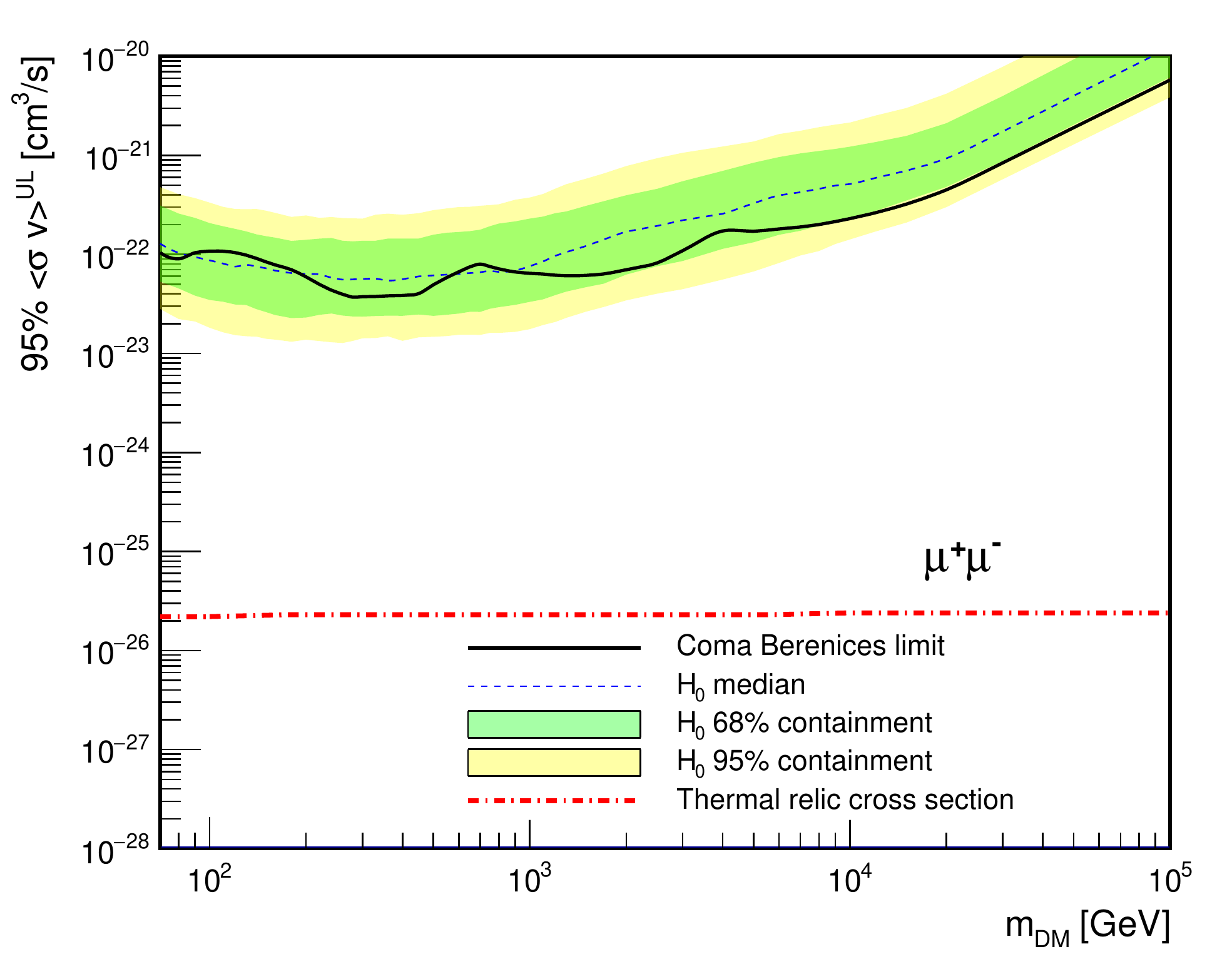}
    \end{subfigure}
    \caption{95\,\% CL ULs for $\langle \sigma_{\mathrm{ann}} v \rangle$ for DM annihilation into $b\bar{b}$, $W^{+}W^{-}$, $\tau^{+}\tau^{-}$ and $\mu^{+}\mu^{-}$ pairs, as representative annihilation channels for both leptonic and hadronic interactions. The black solid line indicates the observed limits obtained for 49.8\,h of Coma Berenices dSph observations, while the blue dashed line is the median of the 300 realizations of the null hypothesis. The green and yellow bands represent the two sided 68\,\% and 95\,\% containment bands respectively. The red dashed line shows the thermal relic cross-section~\cite{2012PhRvD..86b3506S}.}
    \label{fig:limits-Coma}
\end{figure}

The full likelihood method, described in Section~\ref{sec:likelihood-analysis}, was applied to the data, using the $J$-factors reported in Table~\ref{tab:dSphs-list} together with their respective uncertainties and considering single-channel annihilation modes. Note that if the $J$-factor uncertainty is asymmetric, the negative value was adopted as it is the one decreasing our sensitivity to DM signals. The ULs at 95\,\% CL on $\langle \sigma_{\mathrm{ann}} v \rangle$ derived from Equation~\ref{eq:upper-limits} are shown, as a function of the DM mass, in Figure~\ref{fig:limits-Draco} for Draco dSph and in Figure~\ref{fig:limits-Coma} for Coma Berenices dSph. Here, the results are shown only for DM particles annihilating into $b\bar{b}$, $\tau^+\tau^-$, $\mu^+\mu^-$ or $W^+W^-$ pairs, as representative annihilation channels for both leptonic and hadronic interactions. The two-sided 68\,\% and 95\,\% containment bands for the distribution of limits and the median of this distribution, calculated from a sample of 300 simulations of the null hypothesis $\langle \sigma_{\mathrm{ann}} v \rangle = 0$, are also shown. In practice, for each simulation we generated new events for the ON and the OFF regions from the background probability density function, thus assuming no DM signal is present, and computed UL on $\langle \sigma_{\mathrm{ann}} v \rangle$ using exactly the same procedure as for the data. This procedure allowed us to estimate the probability density function for $\langle \sigma_{\mathrm{ann}} v \rangle^{\mathrm{UL}}$ for the null hypothesis case, and double-check that no significant deviations (positive for the case of signal or negative for the case of uncontrolled systematic errors) are present. The results do not show any significant signal related to DM for either of the dSphs, since the achieved limits are within the 68\,\% containment band.

Considering the $b\bar{b}$ and the $\tau^+\tau^-$ channels, the best velocity-averaged cross-section limit for Draco dSph reaches $5.1 \times 10^{-23}$\,cm$^3$/s for a 5\,TeV DM mass and $7.4 \times 10^{-24}$\,cm$^3$/s for a 1.2\,TeV DM mass, respectively. In the case of Coma Berenices dSph, the best limits on $\langle \sigma_{\mathrm{ann}} v \rangle$ for the $b\bar{b}$ and the $\tau^+\tau^-$ channels reach  $ 5.6 \times 10^{-23}$\,cm$^3$/s for a DM mass of 2\,TeV  and $1.1 \times 10^{-23}$\,cm$^3$/s for a DM mass of 0.5\,TeV, respectively. The results obtained are comparable between the two dSphs, due to the similarity of their $J$-factors and exposure time. A comparison of the newly obtained limits with the ones derived from Segue~1 and Ursa Major~II is presented in Section~\ref{sec:combined-limits}.

\section{Ursa Major~II and Segue~1 analyses\label{sec:Ursa-and-Segue-results}}

The published results from Ursa Major~II observations~\cite{2018JCAP...03..009A} were previously obtained using the same approach as the one presented here for Draco and Coma Berenices dSphs. We therefore do not introduce any change either to the Ursa Major~II dataset, consisting of a total of 94.8\,h of good quality data, or its analysis. In order to combine the results obtained from the different targets in a uniform way, we include for the first time in this paper the treatment of the extension of the Segue~1 dSph by means of the \textit{Donut} MC method, as for the other considered targets. The dataset consisting of a total of 157.9\,h of good quality data was therefore left unchanged with respect to the previous publications in~\cite{2014JCAP...02..008A} and~\cite{2016JCAP...02..039M} of the MAGIC data on Segue~1, except for the IRFs that now include the morphology of the target. Also, an updated $J$-factor estimate from~\cite{2015ApJ...801...74G}, whose value is reported in Table~\ref{tab:dSphs-list}, has been adopted as described in Section~\ref{sec:MAGIC-dSphs}. Note that given that the extension of Segue~1 is not much larger than the MAGIC angular resolution (the angular galactocentric distance of the outermost member star is $0.35^\circ$ in~\cite{2015ApJ...801...74G}), the results computed accounting for its extension differ by less than 10\,\% compared to the results derived from a point-like analysis, thus much smaller than the statistical error on the $J$-factor.

\section{Combined limits and discussion\label{sec:combined-limits}}

\begin{figure}
    \hspace{-1cm}
    \begin{tabular}{ccc} 
        \includegraphics[width=0.33\textwidth]{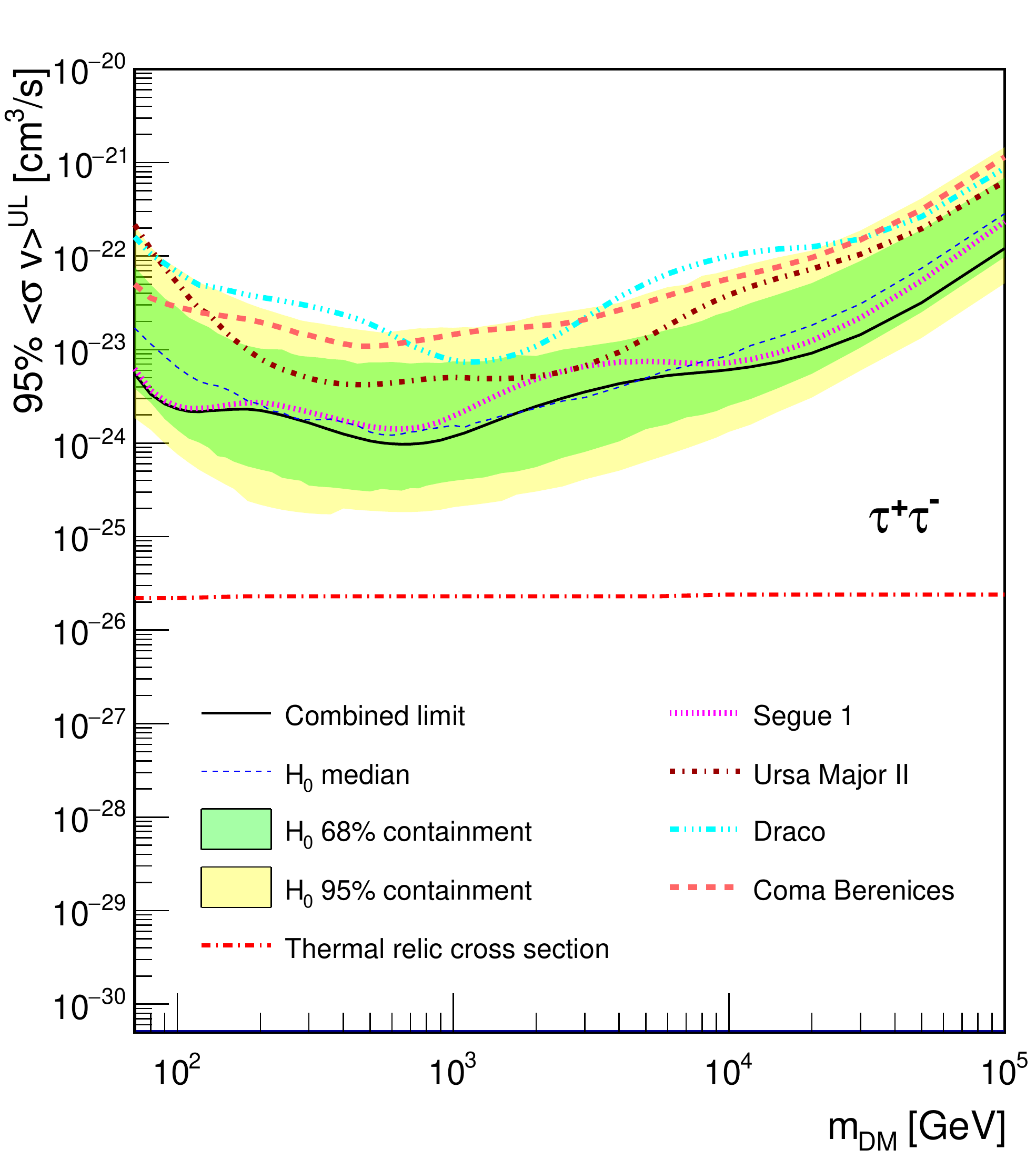} &
        \includegraphics[width=0.33\textwidth]{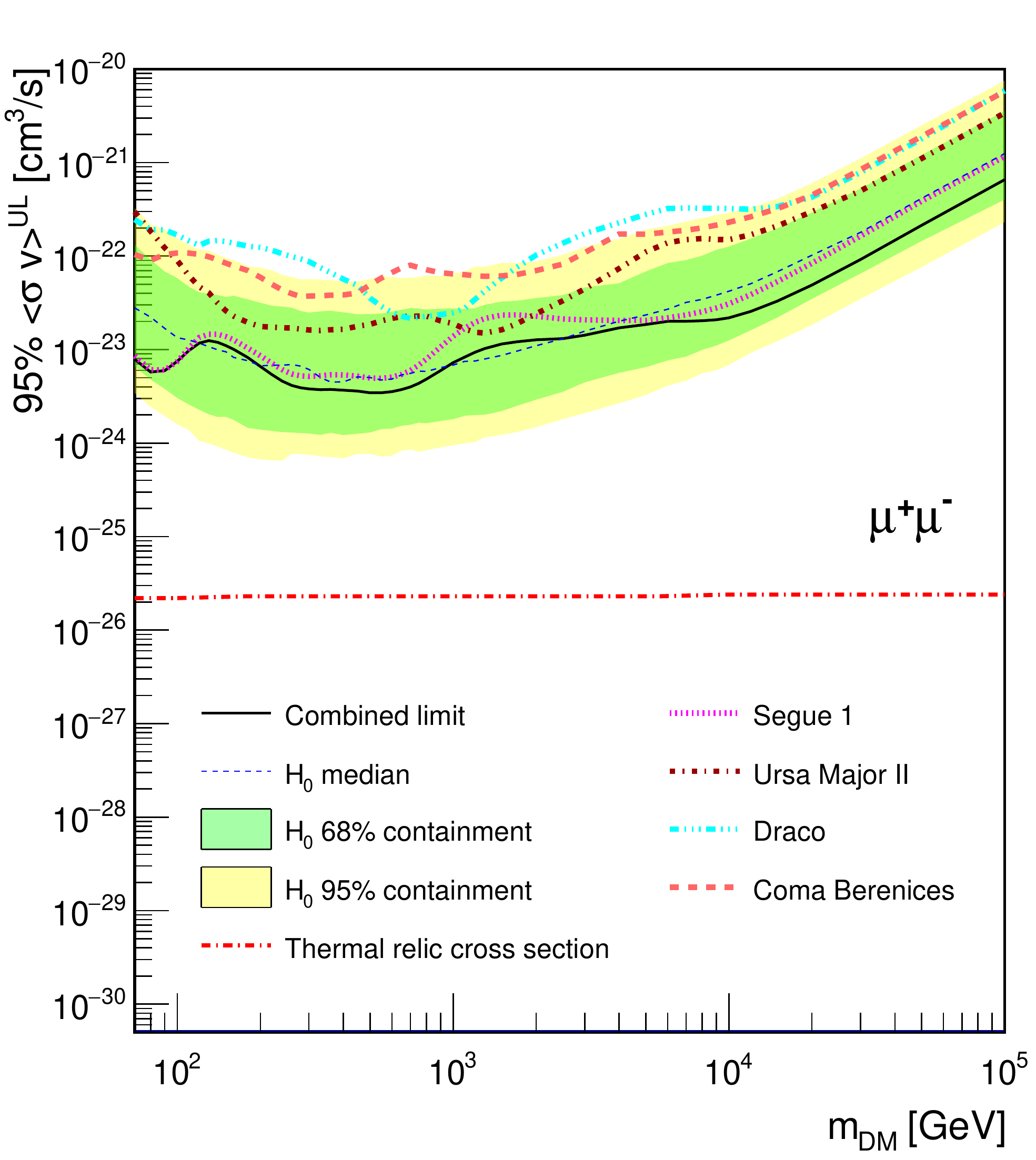} &
        \includegraphics[width=0.33\textwidth]{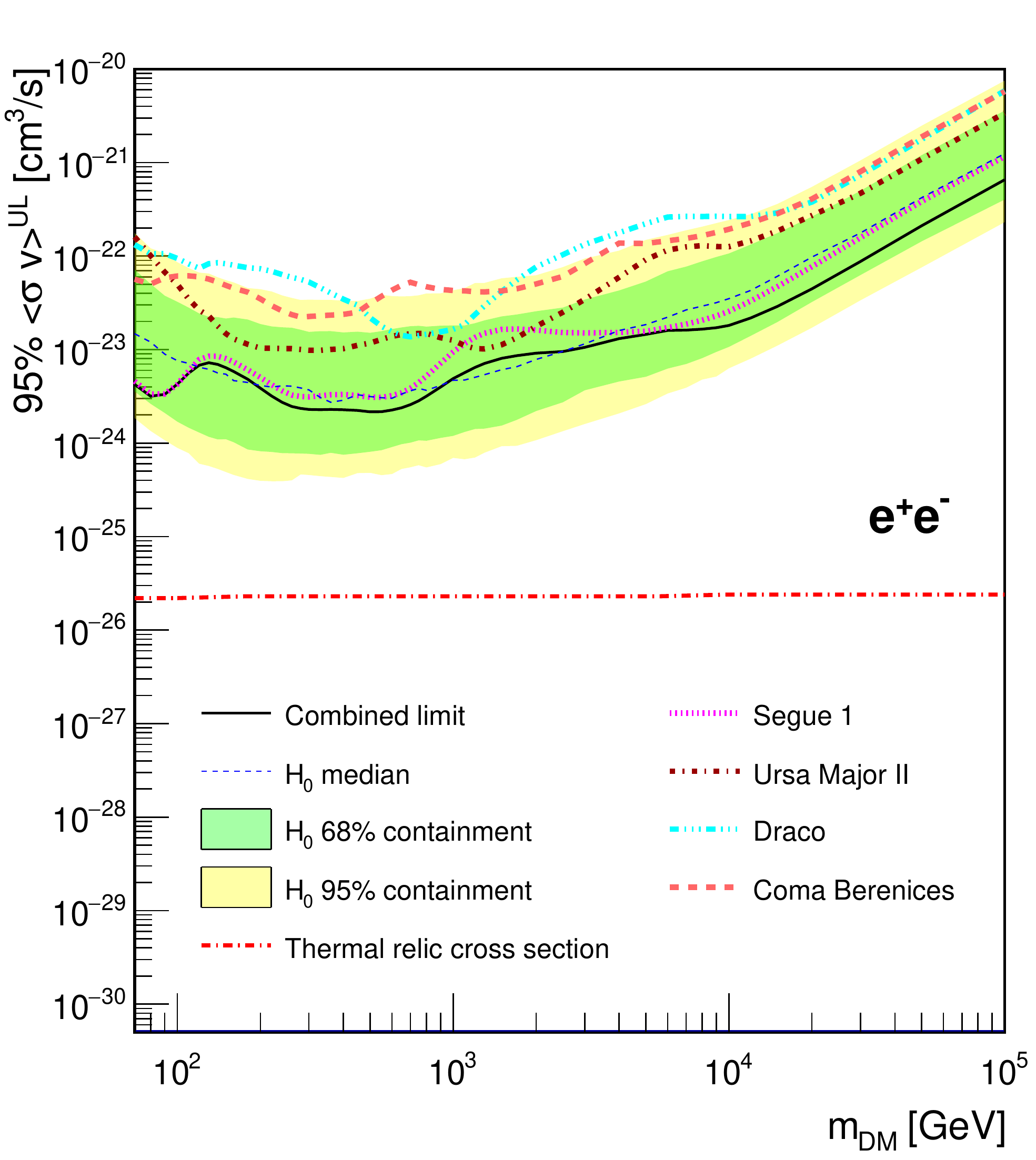} \\
        \includegraphics[width=0.33\textwidth]{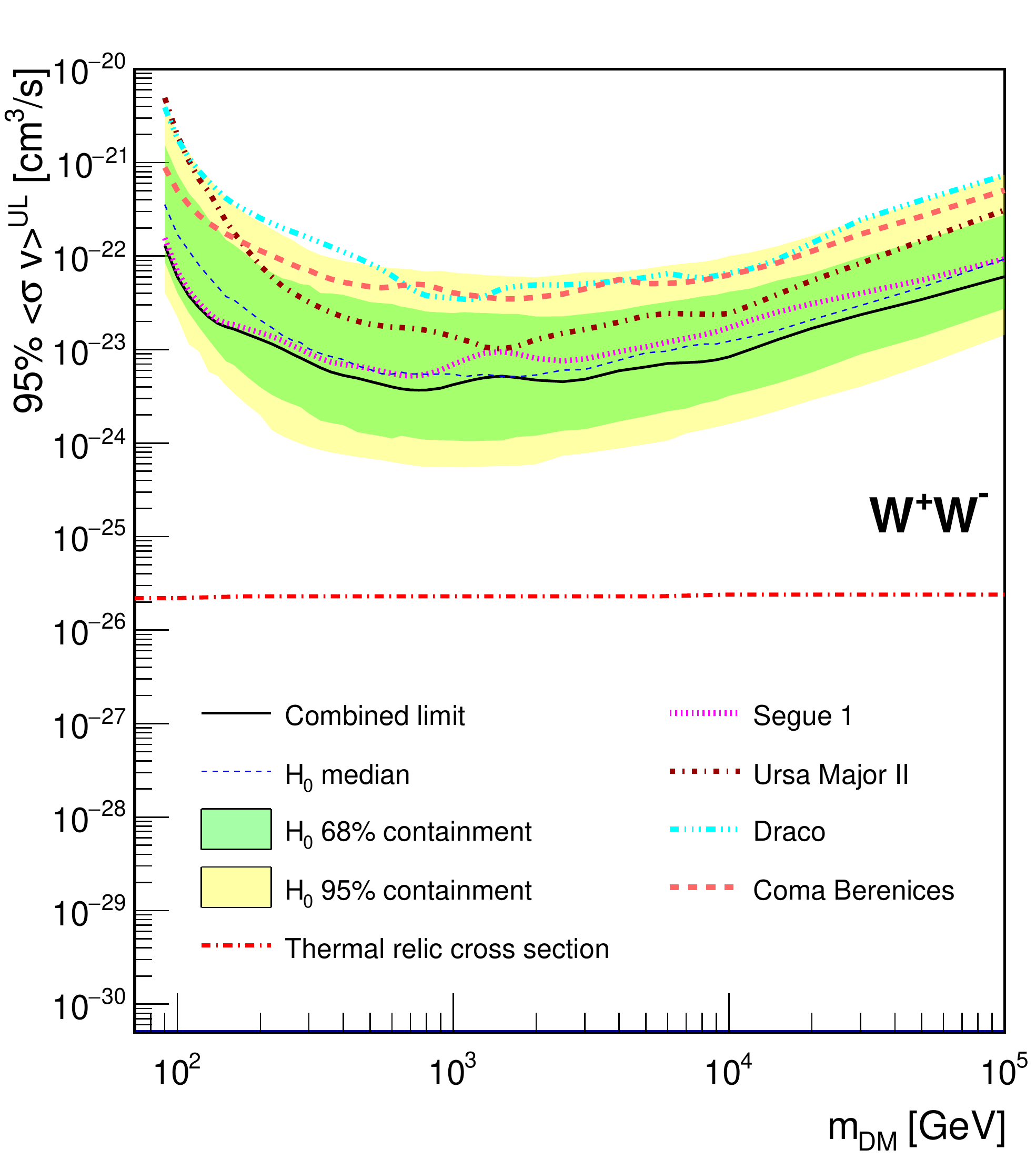} & 
        \includegraphics[width=0.33\textwidth]{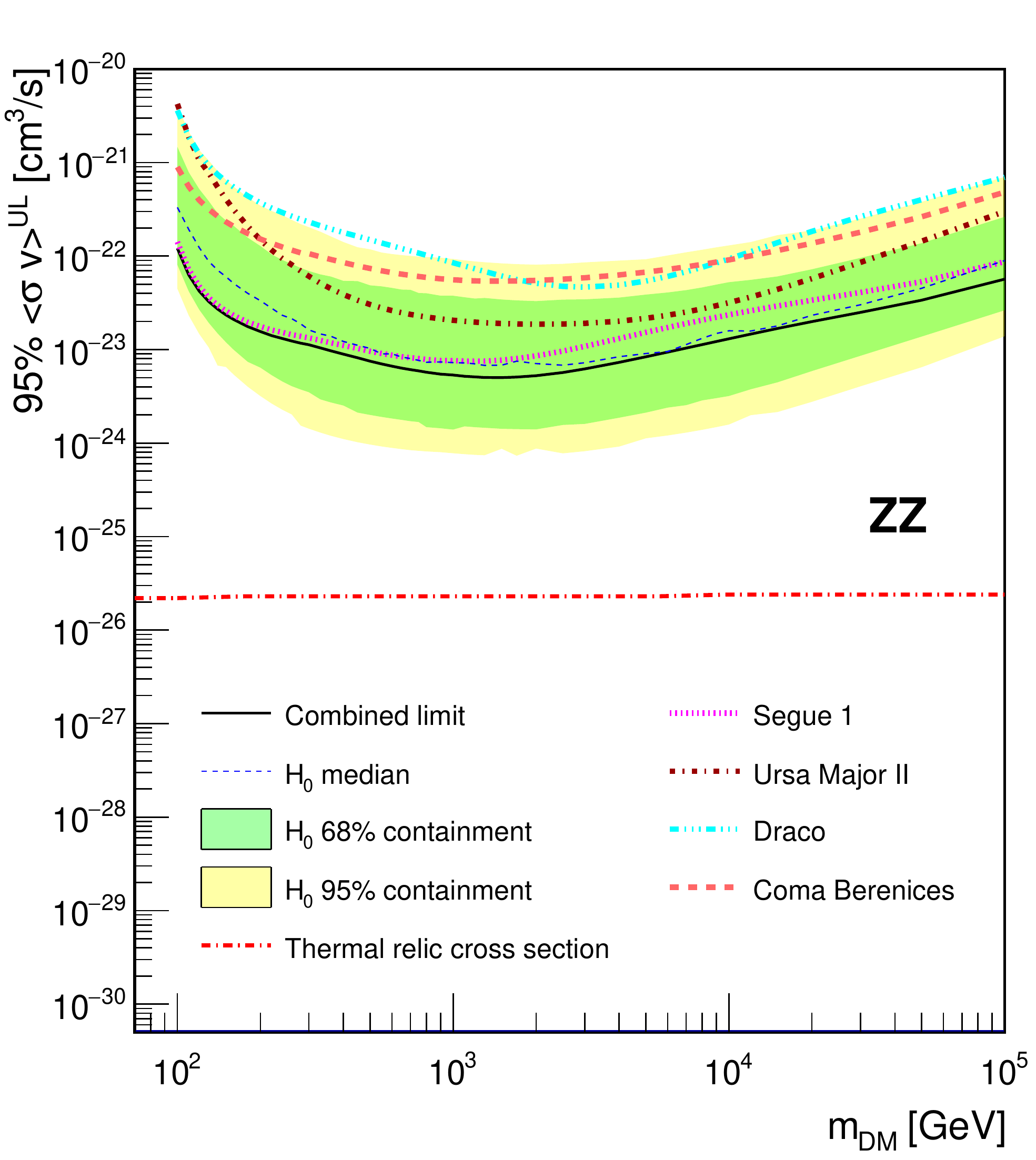} &
        \includegraphics[width=0.33\textwidth]{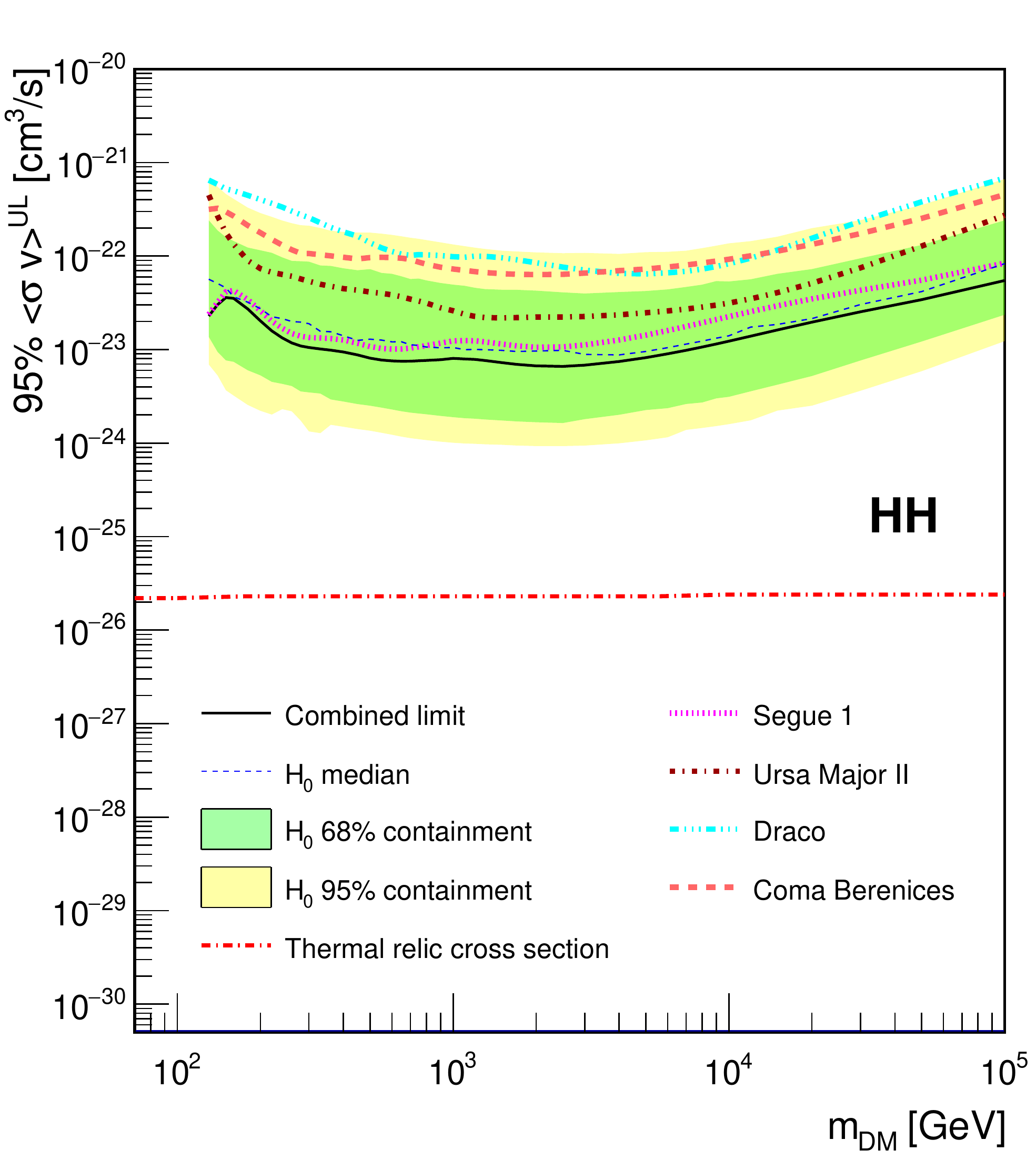} \\
        \includegraphics[width=0.33\textwidth]{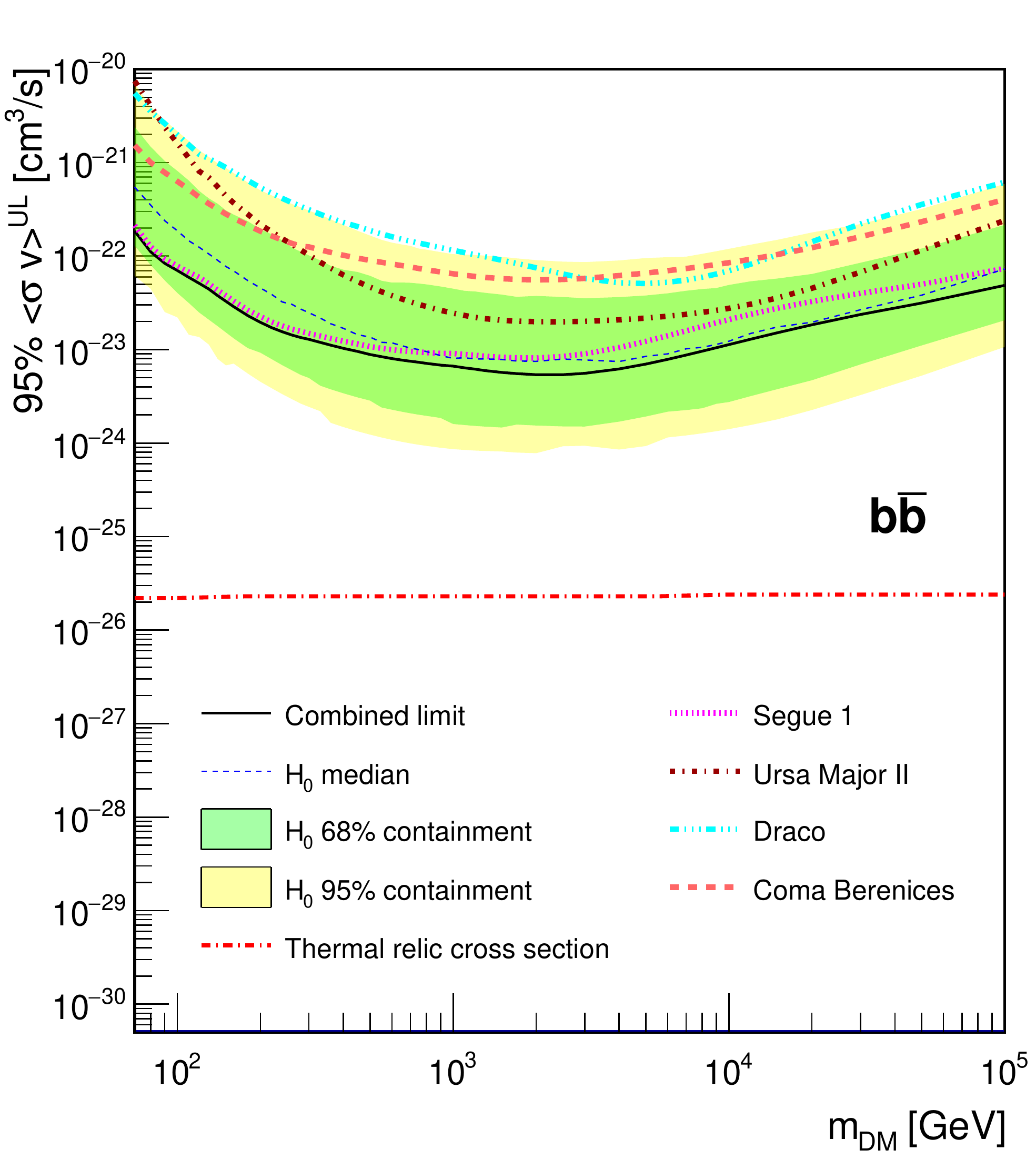} & 
        \includegraphics[width=0.33\textwidth]{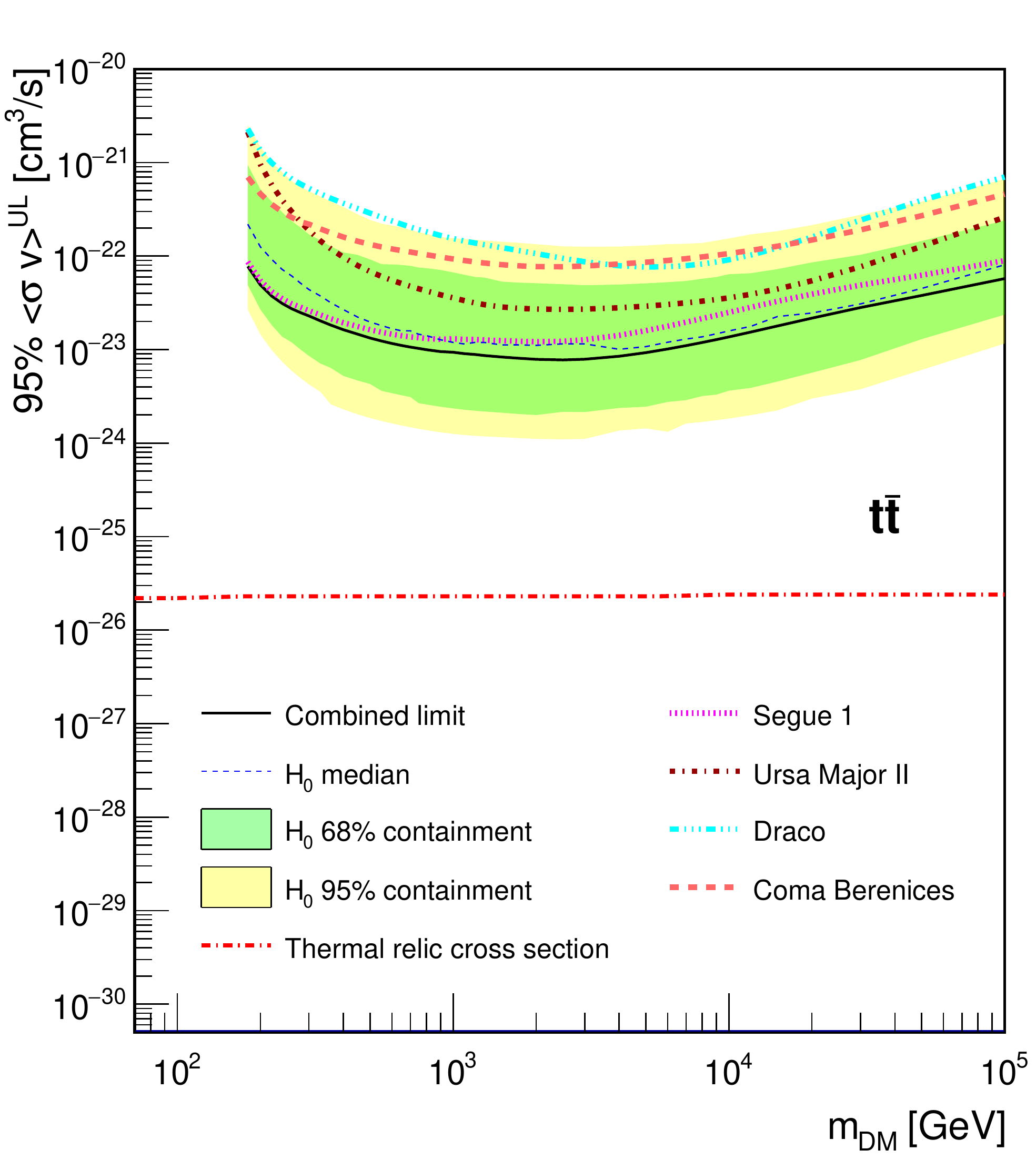} &
        \includegraphics[width=0.33\textwidth]{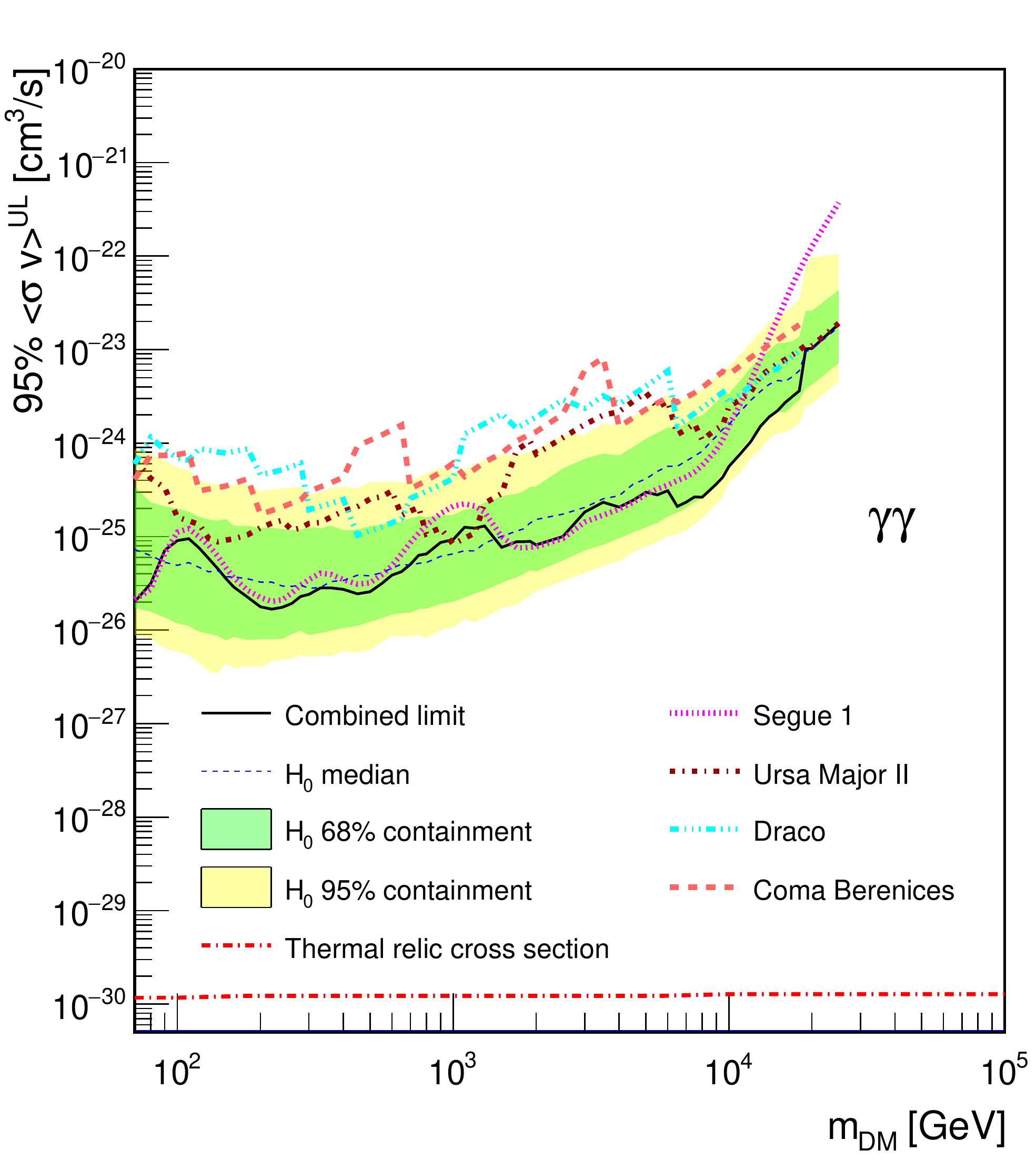} \\
    \end{tabular}
    \caption{95\,\% CL ULs for $\langle \sigma_{\mathrm{ann}} v \rangle$ for DM annihilation into $e^+e^-$, $\mu^+\mu^-$, $\tau^+\tau^-$, $W^+W^-$, $ZZ$, $HH$, $b\bar{b}$, $t\bar{t}$ and $\gamma \gamma$. The black solid line indicates the observed combined limits obtained for 354.3\,h of dSphs observations, while the blue dashed line is the median of the 300 realizations of the null hypothesis. The green and yellow bands represent the two sided 68\,\% and 95\,\% containment bands respectively. The red dashed line shows the thermal relic cross-section~\cite{2012PhRvD..86b3506S}. The one for the $\gamma\gamma$ annihilation channel is multiplied by a factor $\alpha^2$, where $\alpha$ is the fine structure constant (see text for details)}.
    \label{fig:combined-limits}
\end{figure}

We combine the individual datasets in the maximization of a joint likelihood function for all observed targets, observation periods and pointing directions (each described by Equation~\ref{eq:binned-likelihood-formula}), as written in Equation~\ref{eq:combined-likelihood-formula}. The combined limits are therefore computed using a total of 354.3\,h of good quality data. The 95\,\% ULs on the velocity-averaged cross-section $\langle \sigma_{\mathrm{ann}} v \rangle$ are reported in Figure~\ref{fig:combined-limits} for each of the 9 annihilation channels considered. In addition, we show the limits corresponding to each considered dSph. The global limits are mostly within the 68\,\% containment band of the null hypothesis. They are dominated by Segue~1 results, but they are nevertheless more constraining for each channel and for almost every mass. At the lower DM masses (below 1\,TeV) the improvement is marginal, ${\sim}10$\,\% at most, while at higher masses the improvement reaches ${\sim}$40--50\,\% due to the fact that Segue~1 alone limits are less dominant in this regime.

The constraints on $\langle \sigma_{\mathrm{ann}} v \rangle$ from the combined analysis reported here are the most stringent ones obtained with the MAGIC telescopes up to now. A substantial improvement of the limits was achieved by stacking all available targets. Excluding the result obtained for the $\gamma\gamma$ annihilation channel, for which the thermal relic cross-section is multiplied by the fine structure constant squared when assuming 100\,\% branching ratio into photons pairs\footnote{Naively, one would expect this process to happen at a rate of $\alpha^2\langle \sigma_{\mathrm{ann}} v \rangle$, hence suppressing this channel by a factor $\left(\ddfrac{1}{137^2}\right)^{-1} \sim 10^4$~\cite{1994Bergstrom}.}, the closest UL to the thermal relic cross-section is the one relative to $\tau^{+}\tau^{-}$, excluding $\langle \sigma_{\mathrm{ann}} v \rangle$ down to ${\sim}1 \times 10^{-24}$\,cm$^3$/s for DM masses in the TeV range.

It should be remarked that the $J$-factor values used to calculate the above mentioned ULs are affected by target-related systematic uncertainties, such as misidentified foreground interloping stars as described in~\cite{2016MNRAS.462..223B} for the case of Segue~1, and by model-related systematic uncertainties from the fact of having assumed the Navarro-Frenk-White density profile~\cite{1996ApJ...462..563N,1997ApJ...490..493N} over other alternatives such as the Einasto profile~\cite{1965TrAlm...5...87E} or the Burkert profile~\cite{1995ApJ...447L..25B}.

\begin{figure}
    \begin{subfigure}
    \centering
    \includegraphics[width=0.49\textwidth]{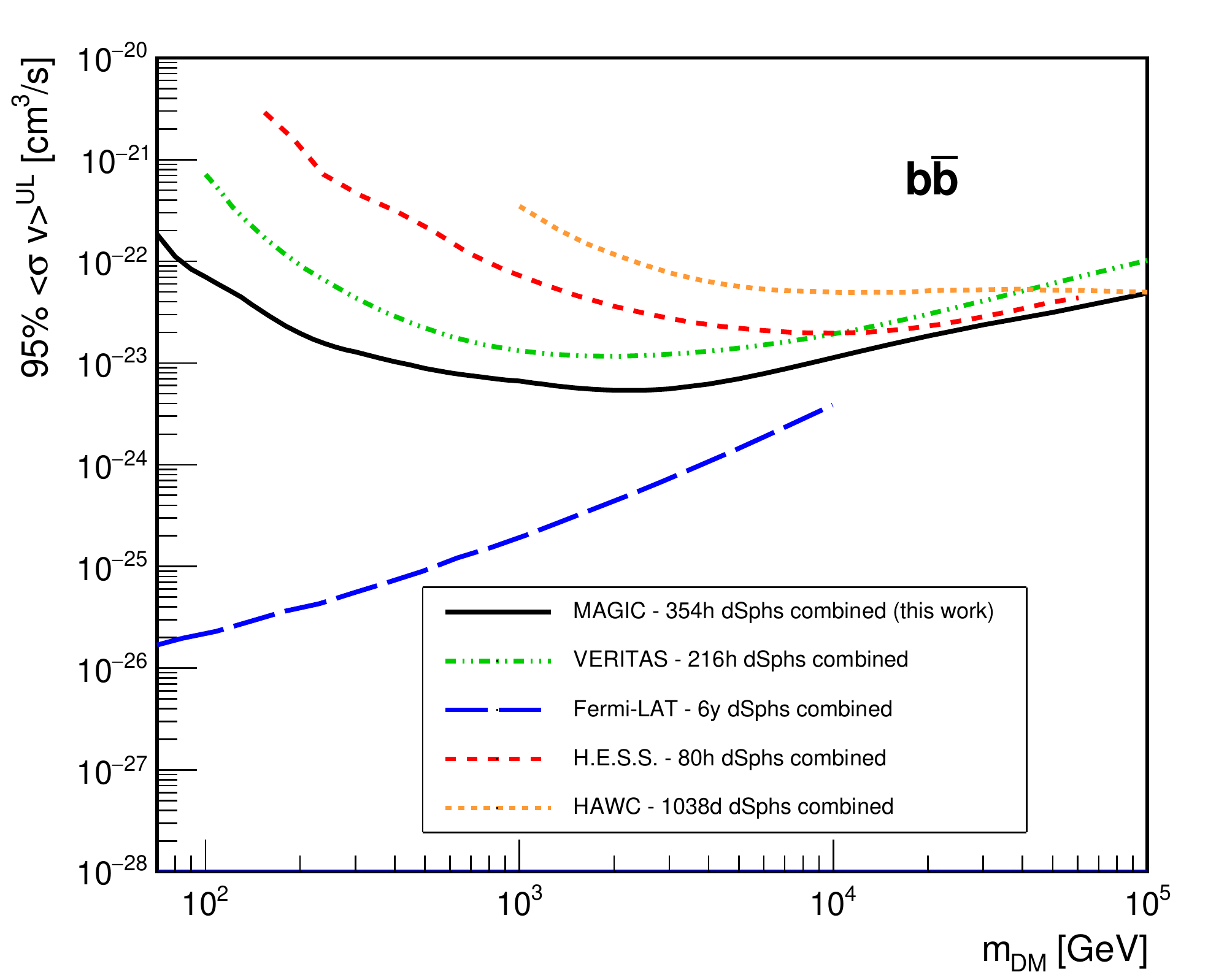}
    \end{subfigure}
    \begin{subfigure}
    \centering
    \includegraphics[width=0.49\textwidth]{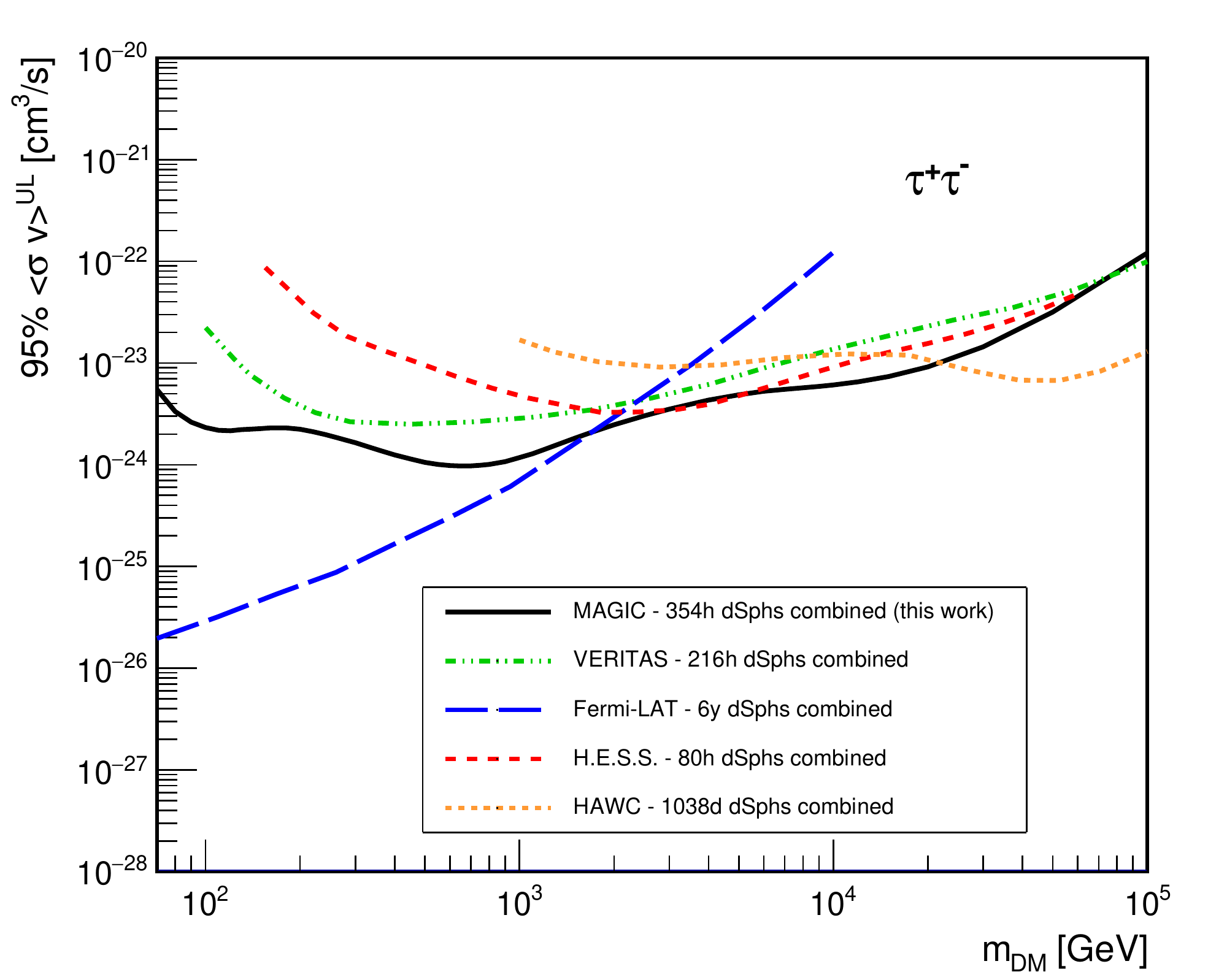}
    \end{subfigure}
    \caption{95\,\% CL ULs on the WIMP velocity-averaged cross-sections for the $b\bar{b}$ (left) and $\tau^+ \tau^-$ (right) channels, from this work (solid black line) and the combined analysis of dSphs from Fermi-LAT~\cite{2015PhRvL.115w1301A} (blue dashed line), VERITAS~\cite{2017PhRvD..95h2001A} (green dashed line), HAWC~\cite{2020PhRvD.101j3001A} (yellow dashed line), and H.E.S.S.~\cite{2020PhRvD.102f2001A} (red dashed line). Note that the three latter results did not include the uncertainties on the $J$-factor in the limits reported here.}
    \label{fig:results-comparison}
\end{figure}

In Figure~\ref{fig:results-comparison} we present the comparison between the combined limits achieved in this work and the ones from dSphs observations by other experiments. MAGIC constraints are the most stringent between a few tens of TeV to 100\,TeV for the $b\bar{b}$ channel, and between few TeV to tens of TeV for the $\tau^{+}\tau^{-}$ channel. Fermi-LAT~\cite{2015PhRvL.115w1301A}, having a better sensitivity at low gamma-ray energies with respect to the other experiments, provides more constraining limits up to TeV DM masses. At higher DM masses, the large duty cycle of the HAWC array sets better $\langle \sigma_{\mathrm{ann}} v \rangle$ ULs for the $\tau^+ \tau^-$ channel~\cite{2020PhRvD.101j3001A}. We remark that these limits remain significantly weaker than the ones claimed by H.E.S.S. on the Galactic Center halo~\cite{2016PhRvL.117k1301A}. However, the Galactic Center halo is affected by large uncertainties, namely by the poorly constrained DM content, not accounted for when producing such limits. On the contrary, the combined limits from dSphs are affected by much smaller uncertainties, thus providing a complementary set of reliable limits.

\section{Summary and conclusions\label{sec:conclusions}}

In this paper, we have presented new results on DM searches obtained by MAGIC from 52.1\,h of observation of the Draco dSph and from 49.5\,h of observation of the Coma Berenices dSph. For both targets we have reported the $\langle \sigma_{\mathrm{ann}} v \rangle$ ULs at the 95\,\% CL for WIMP annihilation in the channels $b\bar{b}$, $\tau^{+}\tau^{-}$, $\mu^{+}\mu^{-}$ and $W^{+}W^{-}$. In order to combine these new results with previous ones in a uniform analysis, we have revised the Segue~1 analysis, taking into account the extension of the source, thanks to the use of the \textit{Donut} MC technique, and considering an updated $J$-factor value: the results previously obtained were not significantly affected. We have then performed a combined analysis of the observations of 4 dSphs for a total of 354.3\,h and have obtained results for the channels $e^+e^-$, $\mu^+\mu^-$, $\tau^+\tau^-$, $W^+W^-$, $ZZ$, $HH$, $b\bar{b}$, $t\bar{t}$ and $\gamma \gamma$. The achieved combined limits from this work are the most stringent in the range from a few TeV to a few tens of TeV among the ones obtained from dSphs observations with IACTs. DM searches combining observations of different targets is now a well established technique in gamma-ray astronomy. It improves the results and strengthens their robustness by averaging out possible systematic uncertainties. The results presented in this paper will be used in a joint analysis of dSphs targets involving different experiments~\cite{2019ICRC...36..539O} that will further maximize the sensitivity of indirect gamma-ray search for DM.

\section*{Acknowledgments}
We would like to thank the Instituto de Astrof\'{\i}sica de Canarias for the excellent working conditions at the Observatorio del Roque de los Muchachos in La Palma. The financial support of the German BMBF, MPG and HGF; the Italian INFN and INAF; the Swiss National Fund SNF; the ERDF under the Spanish Ministerio de Ciencia e Innovaci\'{o}n (MICINN) (PID2019-104114RB-C31, PID2019-104114RB-C32, PID2019-104114RB-C33, PID2019-105510GB-C31, PID2019-107847RB-C41, PID2019-107847RB-C42, PID2019-107988GB-C22); the Indian Department of Atomic Energy; the Japanese ICRR, the University of Tokyo, JSPS, and MEXT; the Bulgarian Ministry of Education and Science, National RI Roadmap Project DO1-400/18.12.2020 and the Academy of Finland grant nr. 320045 is gratefully acknowledged. This work was also supported by the Spanish Centro de Excelencia ``Severo Ochoa'' (SEV-2016-0588, CEX2019-000920-S), the Unidad de Excelencia ``Mar\'{\i}a de Maeztu'' (CEX2019-000918-M, MDM-2015-0509-18-2) and by the CERCA program of the Generalitat de Catalunya; by the Croatian Science Foundation (HrZZ) Project IP-2016-06-9782 and the University of Rijeka Project 13.12.1.3.02; by the DFG Collaborative Research Centers SFB823/C4 and SFB876/C3; the Polish National Research Centre grant UMO-2016/22/M/ST9/00382; and by the Brazilian MCTIC, CNPq and FAPERJ.
This project has received funding from the European Union's Horizon 2020 research and innovation programme under the Marie Sk\l{}odowska-Curie grant agreement No. 754510.

\bibliographystyle{unsrt} 
\bibliography{library} 

\end{document}